\documentclass[twocolumn]{aastex701}

\usepackage{nccmath}
\usepackage{booktabs}
\usepackage{graphicx}	% Including figure files
\usepackage{amsmath,amssymb}	% Advanced maths commands
\usepackage{makecell}

\newcommand{\logM}{\ensuremath{\mathrm{log}_{10} (M_{\star}/\mathrm{M}_{\odot})}}

\newcommand{\pipes}{\texttt{Bagpipes}}
\newcommand{\alfa}{\texttt{alf}$\alpha$}

\usepackage{CJKutf8} %package for Chinese characters
\newcommand{\zh}[1]{\begin{CJK}{UTF8}{bsmi}#1\end{CJK}}

\begin{document}
\title{DeepDive: Tracing the early quenching pathways of massive quiescent galaxies at $z>3$ from their star-formation histories and chemical abundances}
\correspondingauthor{Massissilia L. Hamadouche}\email{mhamadouche@umass.edu}

\author[0000-0001-6763-5551]{Massissilia L. Hamadouche}
\affiliation{Department of Astronomy, University of Massachusetts, Amherst, MA 01003, USA}
\email{mhamadouche@umass.edu}

\author[0000-0001-7160-3632]{Katherine E. Whitaker}
\affiliation{Department of Astronomy, University of Massachusetts, Amherst, MA 01003, USA}
\affiliation{Cosmic Dawn Center (DAWN), Denmark}
\email{}

\author[0000-0001-6477-4011]{Francesco Valentino}
\affiliation{Cosmic Dawn Center (DAWN), Denmark}
\affiliation{DTU Space, Technical University of Denmark, Elektrovej 327, 2800 Kgs. Lyngby, Denmark}
\email{}

\author[0000-0002-0243-6575]{Jacqueline Antwi-Danso}\thanks{Dunlap Fellow}
\affiliation{David A. Dunlap Department of Astronomy \& Astrophysics, University of Toronto, 50 St George Street, Toronto, ON M5S 3H4, Canada}
\affiliation{Dunlap Institute for Astronomy \& Astrophysics, University of Toronto, 50 St George Street, Toronto, ON M5S 3H4, Canada}
\affiliation{Department of Astronomy, University of Massachusetts, Amherst, MA 01003, USA}
\email{}

\author[0000-0002-9453-0381]{Kei Ito}
\affiliation{Cosmic Dawn Center (DAWN), Denmark}
\affiliation{DTU Space, Technical University of Denmark, Elektrovej 327, 2800 Kgs. Lyngby, Denmark}
\email{}

\author[0000-0002-9861-4515]{Aliza Beverage}
\affiliation{Observatories of the Carnegie Institution for Science, 813 Santa Barbara Street, Pasadena, CA 91101, USA}
\email{}

\author[0000-0002-6768-8335]{Pengpei Zhu (\zh{朱芃佩)}}
\affiliation{Cosmic Dawn Center (DAWN), Denmark}
\affiliation{DTU Space, Technical University of Denmark, Elektrovej 327, 2800 Kgs. Lyngby, Denmark}
\email{}

\author[0000-0003-2680-005X]{Gabriel Brammer}
\affiliation{Cosmic Dawn Center (DAWN), Denmark}
\affiliation{Niels Bohr Institute, University of Copenhagen, Jagtvej 128, DK-2200, Copenhagen N, Denmark
}
\email{}

\author[0000-0002-5588-9156]{Vasily Kokorev}
\affiliation{Department of Astronomy, The University of Texas at Austin,
Austin, TX 78712, USA}
\email{}

\author[0000-0002-6220-9104]{Gabriella de Lucia}
\affiliation{INAF - Astronomical Observatory of Trieste, via G.B. Tiepolo 11, I-34143 Trieste, Italy}
\affiliation{IFPU - Institute for Fundamental Physics of the Universe, via Beirut 2, 34151, Trieste, Italy}
\email{}
\author[0000-0003-0215-1104]{William M. Baker}
\affiliation{DARK, Niels Bohr Institute, University of Copenhagen, Jagtvej 155A, DK-2200 Copenhagen, Denmark}
\email{}
\author[0000-0002-5228-2244]{Marion Farcy}
\affiliation{Institute for Physics, Laboratory for Galaxy Evolution and Spectral modelling, Ecole Polytechnique Federale de Lausanne, Observatoire de Sauverny, Chemin Pegasi 51, 1290 Versoix, Switzerland}
\email{}

\author[0000-0002-9656-1800]{Anna Gallazzi}
\affiliation{INAF – Osservatorio Astrofisico di Arcetri, Largo Enrico Fermi 5, 50125 Firenze, Italy}
\email{}
\author[0000-0001-9885-4589]{Steven Gillman}
\affiliation{Cosmic Dawn Center (DAWN), Denmark}
\affiliation{DTU Space, Technical University of Denmark, Elektrovej 327, 2800 Kgs. Lyngby, Denmark}
\email{}
\author[0000-0003-0205-9826]{Rashmi Gottumukkala}
\affiliation{Cosmic Dawn Center (DAWN), Denmark}
\affiliation{Niels Bohr Institute, University of Copenhagen, Jagtvej 128, DK-2200, Copenhagen N, Denmark
}
\email{}
\author[0000-0002-3301-3321]{Michaela Hirschmann}
\affiliation{Institute for Physics, Laboratory for Galaxy Evolution and Spectral modelling, Ecole Polytechnique Federale de Lausanne, Observatoire de Sauverny, Chemin Pegasi 51, 1290 Versoix, Switzerland}
\email{}

\author[0000-0002-8896-6496]{Christian Kragh Jespersen}
\affiliation{Department of Astrophysical Sciences, Princeton University, Princeton, NJ 08544, USA}
\email{}

\author[0000-0003-2918-9890]
{Takumi Kakimoto}
\affiliation{
Department of Astronomical Science, The Graduate University for Advanced Studies, SOKENDAI, 2-21-1 Osawa, Mitaka, Tokyo 181-8588, Japan}
\affiliation{National Astronomical Observatory of Japan, 2-21-1 Osawa, Mitaka, Tokyo 181-8588, Japan}
\email{}

% \author[0000-0002-7598-5292]{Mariko Kubo}
% \affiliation{Department of Physics and Astronomy, School of Science, Kwansei Gakuin University, 2-1 Gakuen, Sanda, Hyogo 669-1337, Japan}
% \email{}

\author[0000-0002-2419-3068]{Minju M. Lee}
\affiliation{Cosmic Dawn Center (DAWN), Denmark}
\affiliation{DTU Space, Technical University of Denmark, Elektrovej 327, 2800 Kgs. Lyngby, Denmark}
\email{}

\author[0000-0003-3228-7264]{Masato Onodera}
\affiliation{
Department of Astronomical Science, The Graduate University for Advanced Studies, SOKENDAI, 2-21-1 Osawa, Mitaka, Tokyo 181-8588, Japan}
\affiliation{National Astronomical Observatory of Japan, 2-21-1 Osawa, Mitaka, Tokyo 181-8588, Japan}
\email{}
\author[0000-0003-4442-2750]{Rhythm Shimakawa}
\affiliation{Waseda Institute for Advanced Study (WIAS), Waseda University, 1-21-1, Nishi-Waseda, Shinjuku, Tokyo 169-0051, Japan}
\email{}
\author[0000-0002-5756-1904]{Masayuki Tanaka}
\affiliation{
Department of Astronomical Science, The Graduate University for Advanced Studies, SOKENDAI, 2-21-1 Osawa, Mitaka, Tokyo 181-8588, Japan}
\affiliation{National Astronomical Observatory of Japan, 2-21-1 Osawa, Mitaka, Tokyo 181-8588, Japan}
\email{}

\author[0000-0003-1614-196X]{John R. Weaver}\thanks{Brinson Prize Fellow}
\affiliation{MIT Kavli Institute for Astrophysics and Space Research, 70 Vassar Street, Cambridge, MA 02139, USA}
\email{}

\author[0000-0002-9665-0440]{Po-Feng Wu}
\affiliation{Graduate Institute of Astrophysics, National Taiwan University,
Taipei 10617, Taiwan}
\email{}

\begin{abstract}
We investigate the chemical abundances and star-formation histories (SFH) of ten massive ($\logM>10.5$) quiescent galaxies at $3<z<4$ using deep, medium-resolution spectroscopic data obtained as part of the \textit{JWST DeepDive} Cycle 2 GO program. Our \textit{DeepDive} sample demonstrates early formation and quenching times inferred from spectro-photometric fitting, with most galaxies having formed 50\% of their stellar mass by $z \sim 5$, and quenching by $z \sim 4$, showing good agreement across the various SFH parameterizations explored in this work. Though they differ slightly between SFH parameterizations, the inferred formation timescales for the {\it DeepDive} sample span both rapid ($\lesssim$ 100 Myr) and more extended ($\gtrsim$ 200 Myr) episodes, corresponding to star formation occurring over a few to several dynamical times given their compact sizes and high densities at $z\sim3-4$. 
On average, massive quiescent galaxies at $3<z<4$ are $\alpha$-enhanced ($\langle [\alpha/\mathrm{Fe}]\rangle$= $0.22^{+0.22}_{-0.17}$), although there is strong diversity ($\sim0.3$ dex in scatter) among individual [$\alpha$/Fe] values. Our results for $\alpha$-enhancement are consistent with lower-redshift studies, implying weak evolution in [$\alpha$/Fe] from $z \sim 4$ to $z\sim 1$. 
The SFH timescales associated with the low [$\alpha$/Fe] measurements suggest longer formation timescales, potentially pointing to earlier enrichment by Type Ia supernovae, or metals preferentially being removed via outflows driven either by powerful early active galactic nuclei or supernovae. Overall, this work represents the first, statistically representative combined study of the star-formation histories and chemical abundances of massive quiescent galaxies at \hbox{$z>3$}.
\end{abstract} 


\section{Introduction} \label{sec:intro}

In the past few years, \textit{JWST} has confirmed the existence of a high number of massive quiescent galaxies (MQGs) at $z>3$ \citep[][]{valentino2023,carnall2023,nanayakkara_2025,merlin2025,stevenson2025}. Given the high quality of the data coupled with the access to near-infrared wavelengths afforded by \textit{JWST}, these galaxies are excellent laboratories to explore the star-formation and chemical-enrichment histories of galaxies. Specifically, the abundance of key elements from spectral absorption features (at $\lambda_{\mathrm{rest}}>4000$\AA, e.g., Mg$_{b}$) can provide effective empirical determinations of the star-formation timescales of massive galaxies. 
% Indeed, these timescales are important indicators of the processes associated with the cessation of star formation. 

Robust measurements of SFH timescales therefore allow us to place strong constraints on the conditions of the early Universe \citep[e.g. see ][]{beverage2025}. Over the past decade, significant effort has been expended into recovering accurate star-formation histories from spectro-photometric fitting of quiescent galaxies at intermediate redshift \citep[using \pipes\, and \texttt{Prospector}, e.g.][$z<2.5$]{carnall2018,leja2019,tacchella_2022b,hamadouche_connection_2023,beverage2024,Slob2024}, extending to $z>3$ with even greater accuracy since the launch of \textit{JWST} \citep[][]{carnall2024,nanayakkara_2025,baker2925flamingojades}.  

Additionally, analyses using large spectroscopic samples of massive quiescent galaxies at $z \sim 1$ from ground-based observations have allowed statistical constraints on the formation times of massive galaxies from star-formation histories and chemical enrichment \citep[see e.g.][]{carnall2019,beverage2021,hamadouche2022,carnall2022,hamadouche_connection_2023,kaushal2024,nersesian2025,gallazzi2025}.  However, detecting absorption features --- vital for robust metallicity measurements --- at $z>1$ from the ground is difficult, and modeling galaxies with young stellar population ages is equally as hard due to the lack of models. Adding further complexity, as age and metallicity indicators trace the average effective SFH --- which includes in-situ and ex-situ star formation --- a galaxy's merger history can make it challenging to interpret formation and quenching timescales. At high redshift, stellar population ages are easier to measure as evolutionary timescales are shorter, and late-stage merger effects (in bringing in ex-situ stars) are minimized, therefore imposing stronger constraints on theoretical models. Extending studies to higher redshifts is therefore crucial to our understanding of galaxy formation histories and evolution.

That said, the degeneracy between metallicity, age, and dust poses a significant problem when trying to understand the properties of these galaxies at high redshift. Reliably differentiating between different star-formation histories is difficult, and inferring the incorrect metallicity for stellar populations with varying chemical abundance patterns can strongly affect their inferred formation times; for example, low inferred metallicities result in higher inferred stellar population ages and star-formation efficiencies at early times \citep[see e.g.][]{carnall2023,carnall2024,degraaff2024,beverage2025}. This raises significant arguments regarding our understanding of $\Lambda$-CDM cosmology and baryon physics at early times \citep[see also][]{glazebrook2024, jespersen2025, jespersen2025a}. It is therefore crucial to reliably measure the true chemical abundance patterns of massive quiescent galaxies at high redshift. 

Locally, the most massive galaxies appear to be rich in elements that are produced in core-collapse supernovae (CCSNe, or Type II SNe) explosions. The abundances of these elements relative to iron are vital in aiding our understanding of a galaxy's formation history, due to the relative timescales on which $\alpha$-elements and iron-peak elements are produced. For example, O is produced on the order of 10 Myr from the onset of star formation, in contrast to Fe-peak elements from SNe Ia, which are also produced on delayed timescales of 1 Gyr from the onset of star formation \citep[see e.g.][]{woosley_weaver_1995,thomas2003,kobayashi2020}.  

This trend also appears to hold out to higher redshifts; a small body of literature exists on chemical abundances (from ground-based measurements) for massive quiescent galaxies at $z\sim1-2$, including single object studies or measurements from stacking \citep[see][]{lonoce2015,onodera2015,kriek2016,beverage2021,beverage2023}. At $z \sim 3.5$, massive star-forming galaxies are found to be $\alpha$-enhanced, with oxygen abundances measured from deep ground-based spectroscopy \citep[from NIRVANDELS][]{stanton2024} yielding $\mathrm{[O/Fe]} = 0.425 \pm 0.026$. This is consistent with values derived for massive quiescent galaxies at $z > 3$ \citep[e.g., $z \simeq 3.2$;][]{carnall2024}, suggesting that quenching may have little impact on chemical abundance ratios.
At high redshift, findings suggest that abundance measurements (specifically [Mg/Fe]) are higher for massive quiescent galaxies that formed earlier in time, consistent with the `downsizing' scenario \citep[][]{kriek_2019,jafariyazani_2020,beverage2025}. 

Unfortunately, spectral modeling of galaxies at higher redshifts based on SSP models --- particularly those with stellar-population \hbox{ages $< 1$ Gyr} --- is extremely difficult; in addition to the chemical abundance features being weaker for these young stellar populations and requiring high SNRs, current versions of stellar population models are tuned towards low-redshift galaxies with stellar population ages of $>1$ Gyr \citep[][]{thomas2003,vazdekis2015, conroy_2018}. These models thus do not span the parameter space required for the much younger ($<1$ Gyr) quiescent galaxy populations that exist at $z > 2.5$ ($\sim 3$ Gyr after the Big Bang). 

Moreover, current stellar population synthesis models integrated into popular SED fitting codes such as \texttt{Prospector} and \texttt{Bagpipes} \citep[e.g. BPASS, BC03 and FSPS,][respectively]{BPASS2018, BC03, FSPS_code} are based on locally-observed solar abundance ratios, which have been proven to not accurately describe the population of quiescent galaxies \citep[even at low redshift, see][]{thomas2003,beverage2025,park2024b,jespersen2025b_IR}. Fortunately, much effort has been exerted into developing $\alpha$-enhanced models that cover a wider range of stellar-population ages \citep[see e.g.][for sMILES, $\alpha$-MC and the single-star $\alpha$-enhanced BPASS models, respectively]{knowles2023,park2024b,byrne2025}. These models are integrated into codes such as \texttt{Prospector} and \alfa\ to obtain more robust ages and metallicities for young, early quiescent galaxies. 

Motivated by these recent observational and technical developments, we leverage the statistical capabilities afforded by \textit{DeepDive}, a \textit{JWST} Cycle 2 spectroscopic program (ID: 3567, PI: F. Valentino) targeting ten quiescent galaxies to obtain deep, high signal-to-noise, high-quality, medium-resolution spectroscopy between $3 < z < 4$ \citep[][]{Ito_2025}.  By combining the results from spectro-photometric fitting and stellar-population fitting using \pipes\ and \alfa, respectively, we obtain robust measurements for stellar mass, abundances ([$\alpha$/Fe], [Fe/H], and total metallicity), as well as formation and quenching timescales. For the first time, we can place important statistical constraints on the star-formation histories and chemical abundances with medium-resolution spectroscopy of the significant population of quiescent galaxies at $z>3$.

The structure of this paper is as follows: we discuss the \textit{DeepDive} observations in Section \ref{sec:data} and the details of our \pipes\ and \alfa\ fitting in Section \ref{sec:methods}. Our main spectro-photometric fitting results are presented in Section \ref{sec:pipes_analysis} and the results obtained from our \alfa\ fitting are presented in \ref{sec:abundances}. We discuss these combined results in Section \ref{sec:discussion} and provide comparisons to relevant literature. Finally, we summarize our conclusions in Section \ref{sec:summary}. Throughout this paper, we quote all magnitudes in  AB magnitudes \citep[][]{okegunn_abmag}, and use a \cite{kroupaimf} initial mass function for all fitting. We assume a $\Lambda-$CDM cosmology with cosmological parameters of \hbox{H$_{0} = 70$ km s$^{-1}$ Mpc$^{-1}$}, $\Omega_{m} = 0.3$ and $\Omega_{\Lambda} = 0.7$. The \cite{knowles2023} SSP models employed in our \alfa\ fitting use the \cite{grevessenoels1993} Solar abundance scale of  $Z_{\odot} = 0.0198$. Although this abundance scale is now regarded as obsolete,  the isochrones \citep[][]{pietrinferni2004,pietrinferni2006} were implemented using this value, and so the SSP models are bound to this value \citep{knowles2023}. For our \pipes\ fitting we assume a the \cite{asplund2009} Solar
abundance of $Z_{\odot} = 0.0142$. 

\begin{deluxetable*}{cccccc}
\tablecolumns{6}
\setlength{\tabcolsep}{3pt}
\tablecaption{Details of the free parameter ranges and priors adopted for our \texttt{Bagpipes} fitting of the \textit{DeepDive} photometry and spectroscopy (see Section \ref{sec:methods}). Priors listed as logarithmic are uniform in log-base-ten of the parameter. \label{tab:bagpipes_model}}
\tablehead{
\colhead{\textbf{Component}} & \colhead{\textbf{Parameter}} & \colhead{\textbf{Symbol / Unit}}  & \colhead{\textbf{Range}} & \colhead{\textbf{Prior}} & \colhead{\textbf{Hyper-parameters}}
}
\startdata
Global & Redshift & $z$ & $z_{\mathrm{spec}}$ & Gaussian & $\mu = z_{\mathrm{spec}}$, $\sigma = 0.1$ \\
\hline
SFH (continuity) & Total stellar mass formed & $M_{\star}/\mathrm{M_{\odot}}$ & (10$^{6}$, 10$^{13}$) & log & \nodata \\
~ & Stellar metallicity & $Z_{\star}/\mathrm{Z_{\odot}}$ & (0.1, 5) & log & \nodata \\
~ & SFR change ($i \rightarrow i + 1$) & $i(\log_{10}\mathrm{SFR})$ & ($-10$, 10) & Student-$t$ & Default as in \cite{leja2019} \\
\hline
Dust & Attenuation at 5500\,{\AA} & $A_V$/mag & (0, 4) & uniform & \nodata \\
~ & Deviation from Calzetti slope & $\delta$ & ($-0.3$, 0.3) & Gaussian & $\mu=0.0$, $\sigma=0.1$ \\
~ & 2175\,{\AA} bump strength & $B$ & (0, 5) & uniform & \nodata \\
% Nebular &  Ionization parameter & log(\textit{U}) & -3 &   & \nodata \\
\hline
Calibration & Zeroth order & $P_0$ & (0.5, 1.5) & Gaussian & $\mu=1.0$, $\sigma=0.25$ \\
~ & First order & $P_1$ & ($-0.5$, 0.5) & Gaussian & $\mu=0.0$, $\sigma=0.25$ \\
~ & Second order & $P_2$ & ($-0.5$, 0.5) & Gaussian & $\mu=0.0$, $\sigma=0.25$ \\
~ & Third order & $P_3$ & ($-0.5$, 0.5) & Gaussian & $\mu=0.0$, $\sigma=0.25$ \\
~ & Fourth order & $P_4$ & ($-0.5$, 0.5) & Gaussian & $\mu=0.0$, $\sigma=0.25$ \\
\hline
Noise & White-noise scaling & $a$ & (0.1, 10) & log & \nodata \\
~ & Correlated noise amplitude & $b/f_{\mathrm{max}}$ & (0.0001, 1) & log & \nodata  \\
~ & Correlation length & $l/\Delta\lambda$ & (0.01, 1) & log & \nodata  \\
\hline
\enddata
\end{deluxetable*}

\section{The \textit{DeepDive} Program} \label{sec:data}

Data were taken from the \textit{JWST} Cycle 2 GO Program (ID: 3567, PI: F. Valentino) targeting massive quiescent galaxies at $z>3$ using the NIRCam and NIRSpec instruments. The main targets for this program are eleven \textit{UVJ}-selected sources with spectroscopic redshifts determined from previous ground-based observations \citep[][]{schreiber2018,tanaka2019,valentino_2020,forrest2020}. Due to a guide star failure, one target was not observed and is therefore excluded from our analysis. 
The \textit{DeepDive} survey selection, observations and reduction are described in detail in \cite{Ito_2025}. 

\subsection{Spectroscopic observations}

Here we briefly summarize the \textit{DeepDive} spectroscopic observations, whereas we refer the reader to \cite{Ito_2025} for a complete overview. Of the ten galaxies targeted in \textit{DeepDive}, two (ID 80 and 111) were observed as part of EXCELS \citep[ID: 3543, PIs: Carnall, Cullen,][]{carnall2024}. Each spectroscopic target received approximately one to three hours of integration time per mask, using the NIRSpec multi-shutter array (MSA) F170LP/G235M medium-resolution grating. Targets are located in the Ultra-Deep Survey (UDS), Cosmological Evolution Survey (COSMOS), Subaru/XMM-Newton Deep Survey (SXDS), XMM-VIDEO and Extended Groth Strip (EGS) fields. All spectra were reduced using MSA\textsc{Exp} \citep[][]{msaexp_brammer_2023,heintz_2025,degraaff2024}. For Version 4 of the reduction, see \cite{valentino_2025}.

\subsection{Imaging}
Imaging was taken for five targets without existing JWST observations in four filters (F150W, F200W, F356W and F444W) with the \textit{JWST} NIRCam instrument. The rest of the targets benefit from existing ancillary \textit{HST} and \textit{JWST} imaging from the CEERS, PRIMER and COSMOS-WEB surveys, all of which are publicly available on the DAWN \textit{JWST} archive \citep[DJA, ][]{valentino2023}. 

We extract robust photometry for our sources from the imaging using \verb\aperpy\\footnote{https://github.com/astrowhit/aperpy \citep[][]{weaver2024} }, following the method described in \cite{weaver2024}. We refer the reader to \cite{Ito_2025} for a detailed description of the photometric extraction and PSF-matched catalogs. Briefly, total fluxes are extracted by flux-correcting photometric measurements made from five circular apertures using a Kron-like ellipse fit to the PSF-matched images. These are then further corrected by adding light in the PSF outside the circularized radius of the ellipse. %tears again 

\section{Methodology} \label{sec:methods}

\begin{figure*}

\includegraphics[width=\textwidth]{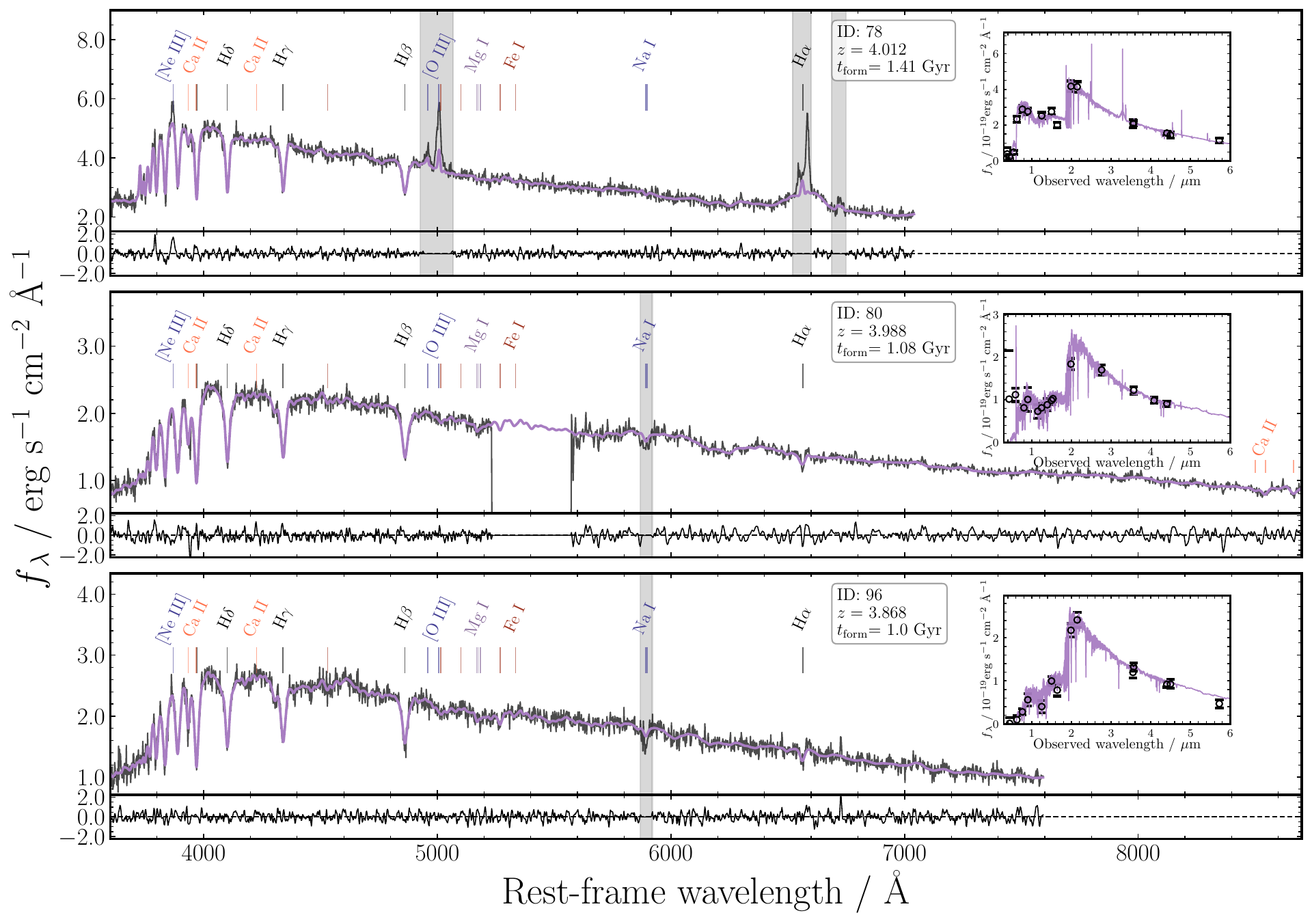}
\caption{Spectra for the three highest redshift galaxies in the \textit{DeepDive} sample (the rest of the galaxies are shown in the Appendix). The fiducial Bagpipes model is over-plotted in purple, and the observed photometry along with the model SED is shown in the inset panel. The shaded gray bands correspond to wavelength regions that were masked during fitting owing to strong emission lines or absorption features indicative of out-flowing gas (e.g., Na D). The top panel shows the fit to DD-78, for which we also include an AGN component, which fits a Gaussian to the broad H$\alpha$ emission (see Section \ref{sec:methods}), while the narrow emission feature is masked due to potential SF contamination. The lower panels show the residuals for each galaxy (where the residuals are $\Delta f_{\lambda}/$errors). }\label{fig:allgals_spec}
\end{figure*}

\subsection{Spectro-photometric fitting}

We use \verb\Bagpipes\ \citep[][]{carnall2018} to simultaneously fit the available photometry and flux-calibrated spectroscopy for our ten main \textit{DeepDive} galaxies using a similar method to that described in \cite{hamadouche_connection_2023}, with some improvements to the star-formation history model.
We use the 2016 version of the \cite{BC03} stellar population models \citep[see][]{bc03_2016} and we assume a \cite{kroupaimf} initial mass function (IMF). Importantly, the BC03 models assume solar abundance ratios based on the \cite{asplund2009} values. In later sections, we describe our independent analysis using \alfa\ \citep[][]{beverage2025} to obtain the chemical abundances of our galaxies using $\alpha$-enhanced simple stellar population models from \cite{knowles2023} to infer empirical star-formation timescales from $\alpha$-element abundances.  The stellar metallicity is varied from $Z_* = 0.1 - 5\ \mathrm{Z}_{\odot}$ using a logarithmic prior. Solar metallicity is assumed to be $Z_{\odot} = 0.0142$ \citep[see][]{asplund2009}. We use the \cite{salim2018} dust attenuation law, which parametrizes the dust-curve shape through a power-law deviation, $\delta$, from the \cite{calzetti2000} law. We use a nebular ionization parameter of $\mathrm{log(}U\mathrm{) = -3}$, which has been shown to be a suitable value for massive quiescent galaxies. We performed additional tests using a uniform prior on $\mathrm{log(}U\mathrm{)}$ between (-4,-1) and find that our final choice does not impact our results with most galaxies having $\mathrm{log}(U) \sim -3$. Additionally, we employ a 4th order spectral calibration polynomial in our fitting to account for issues with flux calibration. However, our results are not dependent on the choice of polynomial order \citep[based on fits with a higher order polynomial calibration, see e.g.][]{carnall2018}. Nebular continuum and emission lines (for H\textsc{ii} regions only, and not including AGN contribution) are modelled using the {\scshape Cloudy} photo-ionization code \citep[][]{ferland2017}, using a method based on that of \cite{byler2017}. The nebular metallicity component and corresponding prior was set to the same range as the stellar metallicity component. Full details of the free parameters and priors used in our fitting are provided in \hbox{Table \ref{tab:bagpipes_model}}.
Additionally, as reported in \cite{Ito_2025}, three galaxies (DD IDs: 78, 170 and 111) appear to possess strong narrow emission in [O III], broad H$\alpha$ emission (DD-78 and DD-111), and broad [He~\textsc{i}] (only object DD-111). During our \pipes\ fitting, we mask the wavelength regions corresponding to these common AGN emission lines for these objects. 
A significant fraction (7/23, 30\%, including secondary targets) of the \textit{DeepDive} galaxies also show evidence for neutral gas outflows via blue-shifted Na \textsc{i} D absorption (P. Zhu et al. in prep.) Recent studies have examined the origin of these features \citep[e.g.][]{belli_2024,davies_2024,deugenio2024,liboni_2025,valentino_2025} in the context of neutral outflows driven by AGN within massive quiescent galaxies; future work will explore this with the \textit{DeepDive} galaxies presented in this paper (P. Zhu et al., in prep.). Following \cite{jafariyazani_2020}, and given the ubiquitous presence of neutral sodium outflows in the \textit{DeepDive} sample, we also mask the neutral sodium absorption feature to avoid complications with metallicity measurements during our fitting. 

\subsubsection{Galaxies with strong AGN emission features}\label{agnemission}
A large fraction of the massive quiescent galaxies observed at $z>3$ show evidence of AGN activity in their spectra \citep[see e.g.][]{stevenson2025,baker2925flamingojades,bugiani2025} 

These particular targets can provide important information on the co-evolution of black holes and quiescent galaxies in the early Universe. However, though interesting, the strong emission features that accompany the presence of an AGN make modeling galaxy SFHs and obtaining robust ages, star-formation rates and metallicities difficult. 
We point out that two galaxies in our sample (DD-78 and DD-111) possess strong, narrow and broad-line AGN emission features and are X-ray detected. We find that these features have a significant effect on our SED modeling; even while masking the regions of strong emission (including wide regions surrounding the features), \pipes\ produces extremely high values for the current star-formation rate for both objects ($> 300$ M$_{\mathrm{\odot}}/\mathrm{yr^{-1}}$) -- unrealistic for these quiescent galaxies at these early times.  For DD-78, we mask the narrow emission lines in the spectrum caused by the AGN (e.g. [O{\scshape iii}] and [O\textsc{ii}], narrow H$\alpha$, [N\textsc{ii}]) and test the addition of the AGN component. This fits Gaussians to the broadened H$\alpha$ and H$\beta$ lines -- as well as a simple power law fit to the continuum -- to account for the AGN contribution (as described in \citealt{carnall2023}). This results in a SFR for the galaxy that is well within our expected values for quiescent galaxies at $z>4$ and in good agreement with the SFR derived from photometry only in \cite{Ito_2025}. We use this additional AGN component in our final fitting for DD-78 only.
%For these tests, we find that they cause a non-negligible fraction of the stellar mass to form at very early times ($z>10$), and extending the star-formation timescale by a few Gyrs, while also decreasing the stellar metallicity to nearly 1 per cent of Solar. 
This highlights the issues of accurately modeling galaxies with strong AGN without a robust AGN model incorporated into our SED fitting methods \citep{jespersen2025b_IR}. We also opt to remove these two galaxies in our \alfa\ modeling, as these issues may have a strong affect on the robustness of our metallicity measurements. 

\subsubsection{Fiducial SFH: the continuity prior}
We use a non-parametric (smooth) continuity prior \citep[][]{leja2019} on the star-formation history which consists of 14 SFR($t$) bins. This SFH is similar to that used in \cite{degraaff2024}, with a slight modification; we adopt the following for the first five time bins (in look-back time) in order to finely sample the most recent star-formation history for each galaxy: 

\begin{align}
\begin{split}
 0 &< t \leq 5\ \mathrm{Myr}, \\ 
 5 & < t \leq 25\ \mathrm{Myr}, \\
 25 &< t \leq 50\ \mathrm{Myr}, \\
 50 &< t \leq 75\ \mathrm{Myr}, \\
 75 &< t \leq 100\ \mathrm{Myr}. \\
\end{split}
\end{align}

The last nine bins are equal-width bins (in log-space) to the age of the Universe at the observed redshift ($z_{\mathrm{zspec}}$) for each galaxy ($t_\mathrm{{obs}}$).

%add details here about the SFH prior choices for the other models iyer, double-power law, and bursty models. 
\begin{deluxetable*}{ccrrccccc}
\setlength{\tabcolsep}{5pt}
\tablecaption{Derived properties from the \texttt{Bagpipes} and \texttt{alf}$\alpha$ (using the sMILES models, see \citealt{knowles2023}) fits for the \textit{DeepDive} galaxies, in order of increasing redshift. $t_{\mathrm{form}}$ is the formation time, defined as the age of the Universe at which the galaxy formed half its stellar mass, and $t_{\mathrm{quench}}$ is the quenching time, defined as the age of the Universe at which the normalized SFR of the galaxy falls below 0.1 \citep[see][]{carnall2018}, obtained from \pipes. We quote the uncertainties for the 16th and 84th percentiles for each quantity.
}
\label{tab:results_table1}
\tablecolumns{9}
\tablehead{
\colhead{DD-ID} & \colhead{ID$^{\ddagger}$} & \colhead{RA} & \colhead{Dec} &
\colhead{$z_{\mathrm{spec}}$} & 
\colhead{$\log_{10}(M_{\star}/M_{\odot})$}  &
\colhead{$\mathrm{log_{10}}(Z_{\star}/Z_{\odot})$} & 
\colhead{$t_{\mathrm{form}}$ / Gyr} & 
\colhead{$t_{\mathrm{quench}}$ / Gyr}} 
\startdata
78$^{\dagger}$ & 262 & 34.298697 & -4.989901 & 4.0117 & $10.94_{-0.03}^{+0.02}$ & $-0.39_{-0.02}^{+0.04}$ & 
$1.41^{+0.01}_{-0.02}$ & \nodata  \\
80$^{\star}$ & 6496 & 34.340358 & -5.241255 & 3.9879 & 
$10.92_{-0.02}^{+0.03}$ & $0.35_{-0.05}^{+0.05}$ &
$1.08^{+0.04}_{-0.04}$ & $1.37^{+0.07}_{-0.05}$  \\
96 & 695 & 34.756280 & -5.308090 & 3.8676 & $10.98_{-0.03}^{+0.01}$ &  $0.07_{-0.13}^{+0.10}$ & 
$1.00_{-0.07}^{+0.04}$ & $1.21_{-0.00}^{+0.00}$  \\
106 & 11245 &149.932856 & 2.123422 & 3.8307 & $11.21_{-0.02}^{+0.03}$ & $-0.34_{-0.05}^{+0.04}$ & 
$1.21^{+0.01}_{-0.03}$ & $1.32_{-0.00}^{+0.12}$ \\
111$^{\dagger}$$^{\star}$ & 3561 & 34.289452 & -5.269803 & 3.7976 & $10.78_{-0.04}^{+0.05}$ & $-0.51_{-0.11}^{+0.29}$ & 
$1.09^{+0.09}_{-0.08}$ & \nodata  \\
115 & 1255 & 149.419594  & 2.007526 & 3.7766 & $10.77_{-0.02}^{+0.02}$ & $0.43_{-0.03}^{+0.03}$ & $1.46^{+0.01}_{-0.01}$ & \nodata  \\
134 & 54459 &150.061466 & 2.378713& 3.7130 & $11.06_{-0.02}^{+0.02}$ & $0.40_{-0.03}^{+0.04}$  & $1.29^{+0.03}_{-0.02}$ & $1.38_{-0.00}^{+0.00}$  \\
170 & 4084  & 36.733541 & -4.536583 & 3.4894 & $11.25_{-0.02}^{+0.03}$ & $0.06_{-0.13}^{+0.12}$ & $1.17^{+0.07}_{-0.11}$ & $1.57_{-0.07}^{+0.06}$  \\
179 & 3386 & 34.386994  & -5.482689 & 3.4472 & $11.49_{-0.01}^{+0.03}$ & $0.32_{-0.09}^{+0.03}$ & $1.36^{+0.04}_{-0.08}$ & $1.52_{-0.00}^{+0.00}$  \\
196 & 61168 &214.866046 & 52.884258 & 3.4332 & $10.94_{-0.02}^{+0.02}$ & $0.03_{-0.06}^{+0.07}$ &  $1.40^{+0.02}_{-0.03}$ & $1.53_{-0.00}^{+0.00}$ \\
\enddata
\tablecomments{\\ $^{\ddagger}$Source ID from previously reported ground-based observations.\\ $^{\dagger}$Removed from the \texttt{alf}$\alpha$ fitting due to significant AGN emission features in its spectrum.\\$^{\star}$Observed as part of the \textit{JWST} EXCELS program \citep[][]{carnall2024}.}
\end{deluxetable*}

\begin{figure*}
\centering\includegraphics[width=\linewidth]{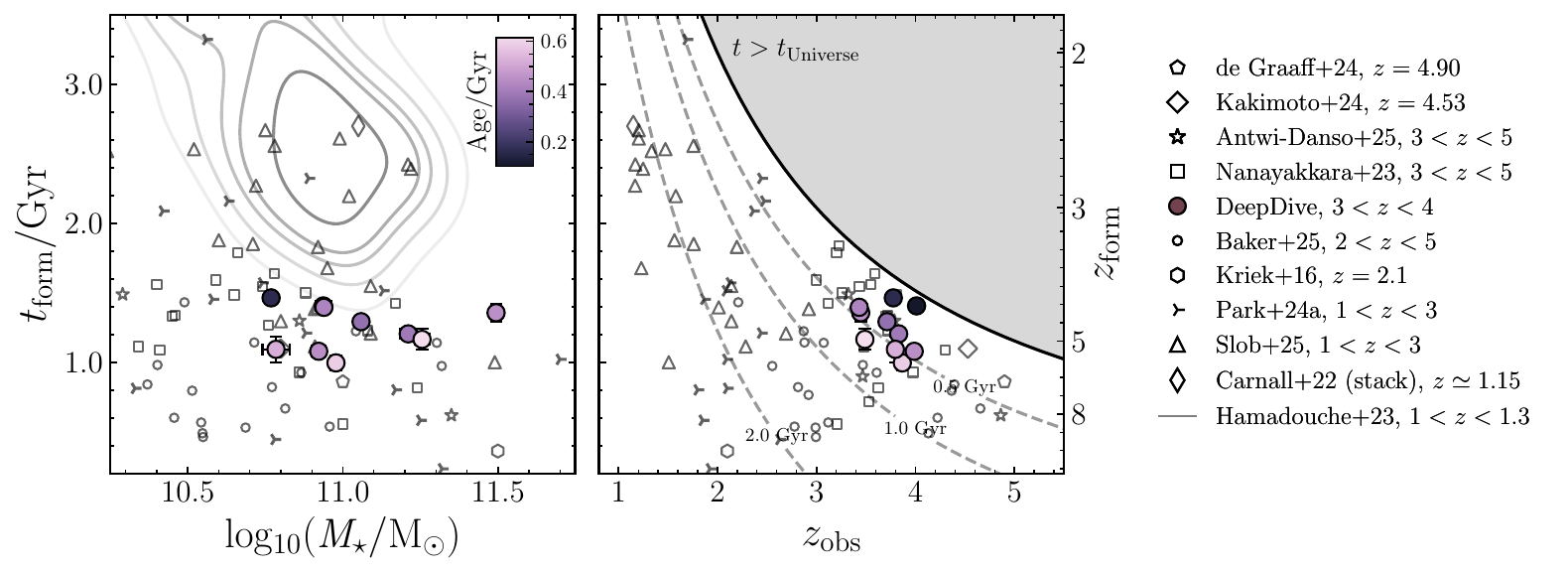}
    \caption{\textit{Left:} Formation time versus galaxy stellar mass for all galaxies in our \textit{DeepDive} sample, alongside a (non-comprehensive, see text for more) number of ground- and space-based literature results at $z>1$. The gray contours represent the massive quiescent galaxies from the VANDELS survey at $1<z<1.3$, presented in \cite{hamadouche_connection_2023}. The \textit{DeepDive} galaxies are color-coded by the mass-weighted age derived from \pipes. \textit{Right:} Formation time versus redshift of observation for spectroscopically confirmed high-redshift quiescent galaxies (up to $ z \sim 5$).  Dashed lines are lines of constant age at 0.5 Gyr, 1.0 Gyr and 2.0 Gyr and the solid black line indicates the age of the Universe as a function of redshift (the shaded gray region is therefore $t>t_{\mathrm{ Universe}}$). }
\label{fig:tform_zobs}
\end{figure*}

\subsection{Priors for other SFH models}
In Section \ref{sec:sfh_comps}, we also show comparisons of our galaxies to three other SFH models commonly employed in the literature -- the parametric double-power law \citep[e.g.][]{carnall2018, hamadouche_connection_2023}, a flexible continuity SFH with a bursty prior \citep[see][]{tacchella_2022b} to account for more variable star formation at early times, and the non-parametric \texttt{DenseBasis} \citep{iyer2019} star-formation history. 

During the three other SFH runs, we use the same priors for redshift, dust, calibration and noise models as our fiducial continuity model. For the bursty SFH prior, we implement the model described in \cite{tacchella_2022b} by increasing the Student-\textit{t} SFR prior from $\sigma = 0.3$ to $\sigma = 1$. This effectively allows for more extreme variation in SFR between neighboring bins compared to our fiducial model. The \texttt{DenseBasis} \citep{iyer2019} SFH model is incorporated into \texttt{Bagpipes} fully and is separate from the FSPS-based SFH model that \texttt{DenseBasis} implements. We use 4 SFR(\textit{t}) bins for this SFH model for better comparison with the literature; additionally, in \cite{iyer2019}, the authors find that higher numbers of time bins produce weaker constraints on the SFH. Interestingly, when we set the Dirichlet prior to $\alpha = 3$ for all bins, we find that it produces a tail of early SFR (also seen when increasing the number of bins). Setting the Dirichlet parameter to the same value for all bins causes equal amounts of stellar mass to be produced in each of the time bins, possibly contributing to this early tail of SFR in our galaxies. We therefore set the Dirichlet parameter for the SFH model to $\alpha = 30$ for the most recent time bin --- in a similar manner as the fine binning we employ for our fiducial model. We find this approach to be suitable in not producing overly sharp star-formation histories \citep[see also][]{harvey2025}, and reducing the amount of stellar mass produced early in time, and bringing the formation times and timescales more in line with the other SFH parameterizations used. Finally, we use a double-power law star-formation history that incorporates two separate power law slopes, where $\alpha$ is the rising slope and $\beta$ is the falling slope, and $\tau$ is the turnover time.

\begin{figure*}
\centering
    \includegraphics[width=\linewidth]{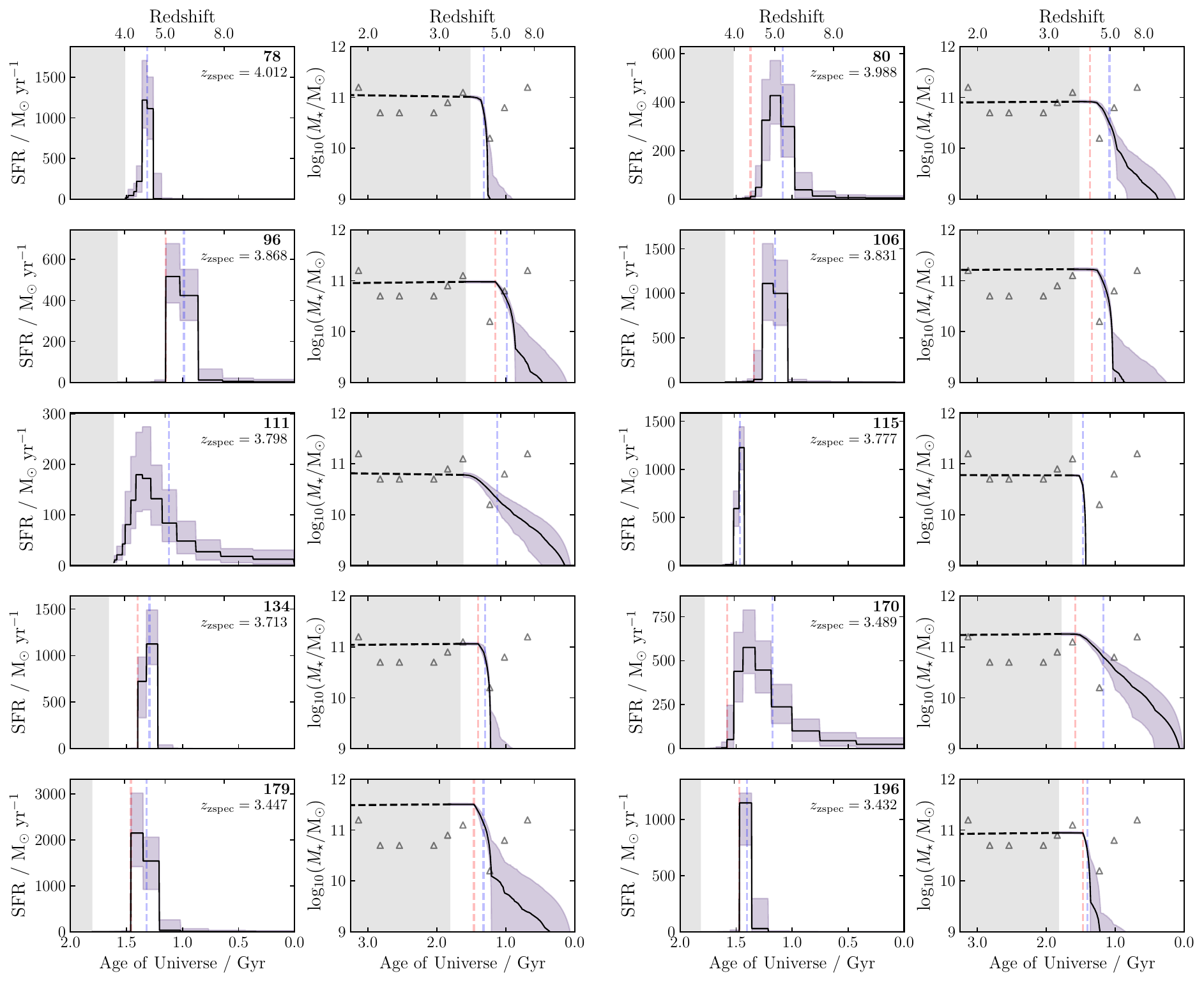
    }
    \caption{The star-formation and cumulative stellar-mass histories for the primary \textit{DeepDive} targets. Horizontal dashed lines are extrapolated values in order to compare the stellar masses to the SUSPENSE galaxies, shown as the unfilled triangles in the cumulative mass history subplots \citep[see][]{Slob2024,beverage2025}. For each column, the blue and red dashed lines correspond to the formation and quenching times obtained from our fiducial \texttt{Bagpipes} run (see Section \ref{sec:methods} for information), respectively. Some galaxies do not have lines corresponding to quenching time, due to the definition of quenched in \pipes; this is explained in further detail in Section \ref{sec:pipes_analysis}. The shaded region in each subplot corresponds to the age of the Universe at the observed redshift.}
    \label{fig:sfhs}
\end{figure*}

\section{Results} 
\label{sec:pipes_analysis}
\begin{figure*}
\centering
\includegraphics[width=\linewidth]{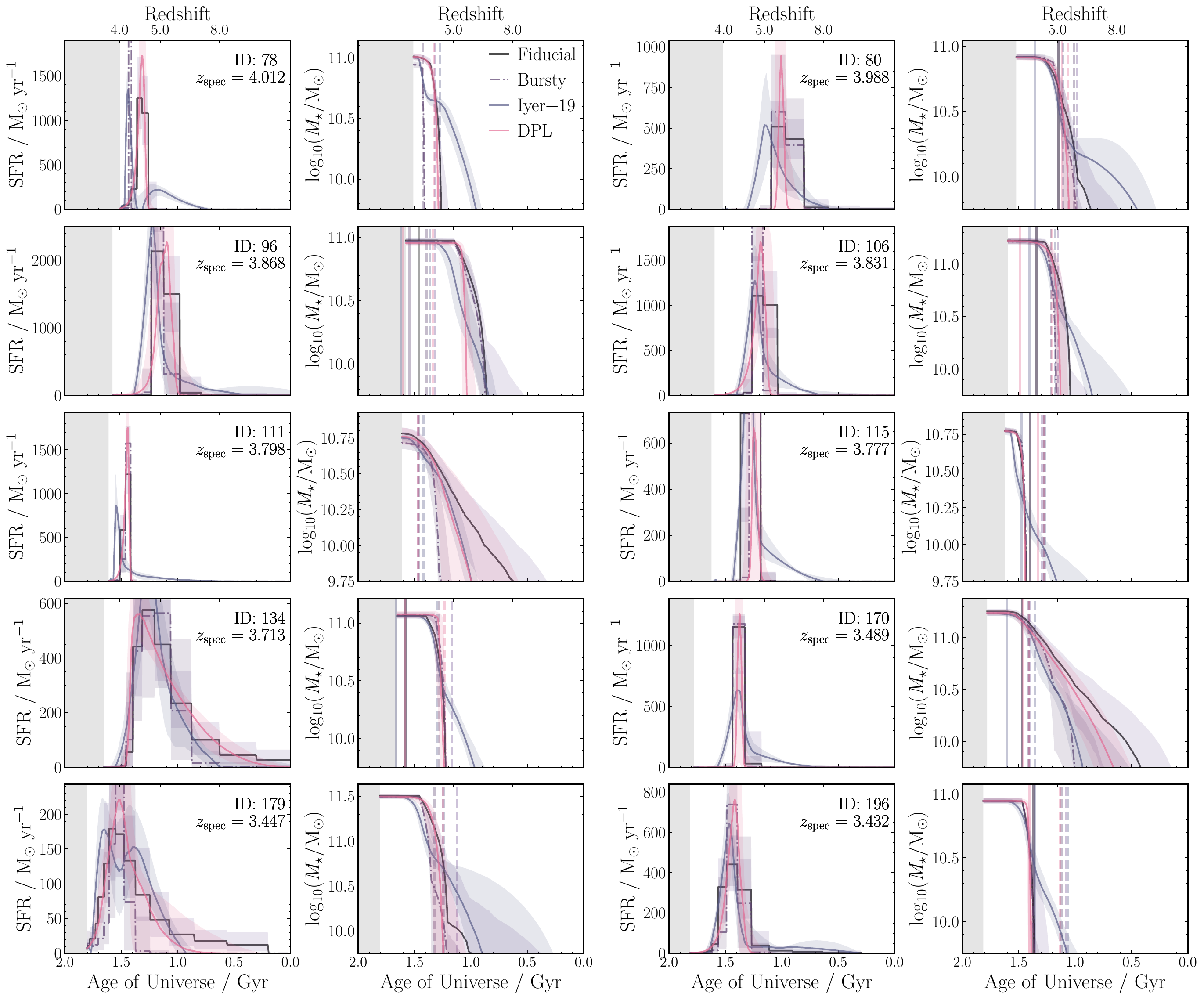}
    \caption{We show a comparison between different star-formation histories with \pipes\ for all of the galaxies in DeepDive (with our fiducial SFH model in black, as in Fig. \ref{fig:sfhs}). Dashed vertical lines represent $t_{\mathrm{form}}$ values for each of the SFHs and solid lines represent the $t_{\mathrm{quench}}$ value (for galaxies that have quenching times from \pipes). The different SFH parameterizations are mostly in good agreement with each other, with only the \texttt{DenseBasis} \citep[][]{iyer2019} tending to produce earlier formation times for some galaxies than the double-power law, bursty or fiducial SFHs. }
\label{fig:sfh_comp_96}
\end{figure*}
\begin{figure*}
\centering
\includegraphics[width = \linewidth]{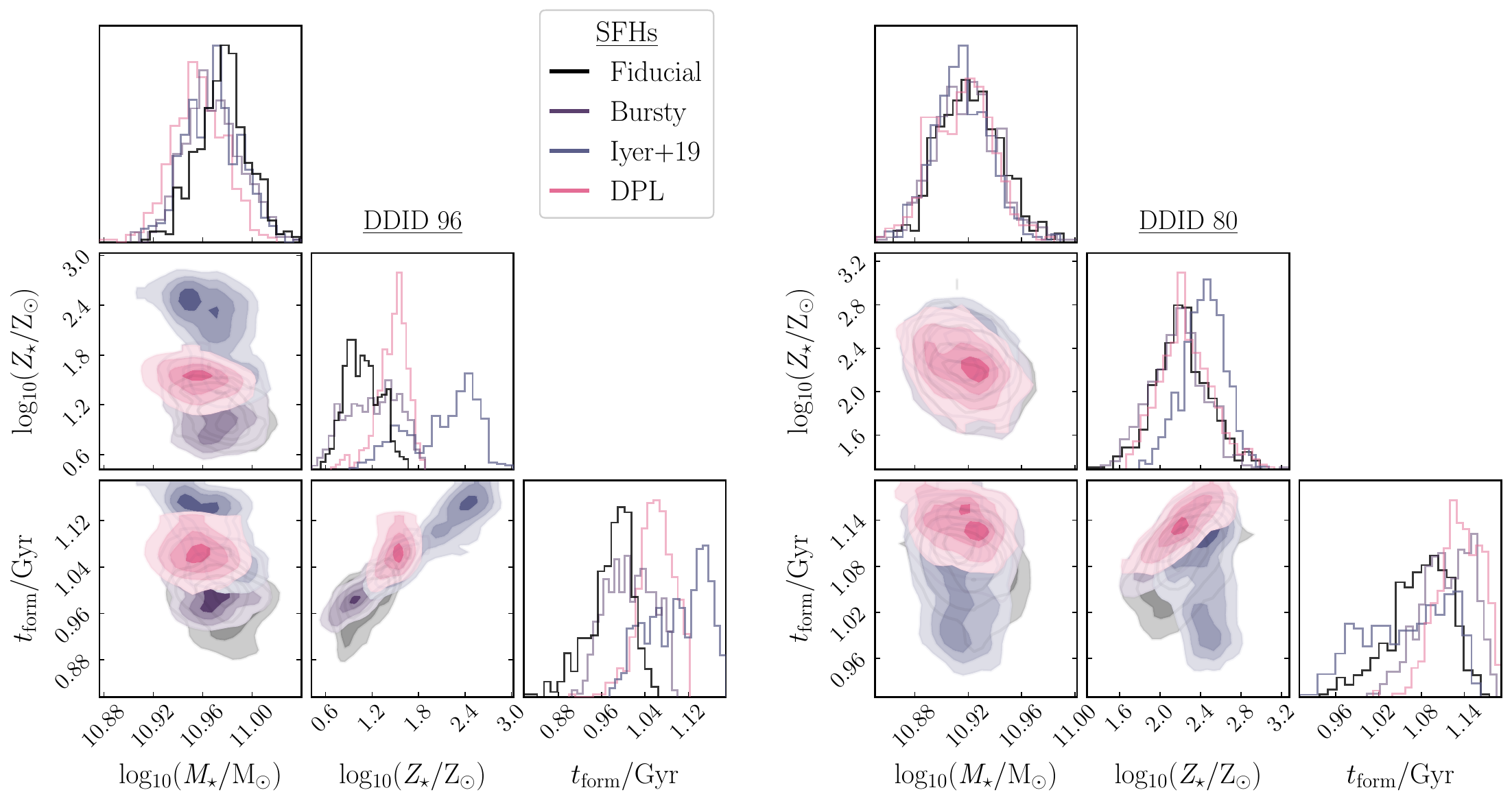}
\caption{Example corner plots for two of the galaxies in the DeepDive sample (DD-96 and DD-80), demonstrating the differences between the sample distributions of each star-formation history prescription for stellar mass, metallicity and formation time. We see that there is very good agreement between SFH models for stellar mass; however, formation times for the double-power law, bursty and \texttt{DenseBasis} SFHs are slightly offset to later times than our fiducial SFH model.}
\label{fig:corners}
\end{figure*}

\subsection{Results from \pipes\ fitting}

We fit our the combined spectroscopy and photometry for the ten \textit{DeepDive} galaxies using \texttt{Bagpipes} and obtain measurements for stellar mass, star-formation rate, metallicity and formation ($t_{\mathrm{form}}$) and quenching times ($t_{\mathrm{quench}}$). The best-fitting results are presented in Table \ref{tab:results_table}. The  stellar masses obtained from our \texttt{Bagpipes} fitting are in excellent agreement (consistent with 1$\sigma$) with those reported in \cite{Ito_2025} from photometric fitting only. In Fig. \ref{fig:allgals_spec}, we present spectra for three of our \textit{DeepDive} galaxies, over-plotted with the best-fitting \texttt{Bagpipes} model in pink and common absorption/emission lines. In each panel, the inset figure displays the photometric data points and the best-fitting SED model. The presence of deep, higher-order Balmer absorption lines in our spectra, combined with the shallow Ca K absorption feature (except for those galaxies with evidence of excess absorption from neutral gas), point to the young, post-starburst nature of quiescent galaxies in the early Universe. 

\subsubsection{Formation and quenching times of massive galaxies}

In the left-hand panel of Fig. \ref{fig:tform_zobs}, we show the formation times versus stellar mass for the ten galaxies in our \textit{DeepDive} sample. The formation time, $t_{\mathrm{form}}$ is defined as the age of the Universe at which fifty percent of the final stellar mass is formed, equivalent to $t_{50}$ in other work \citep[][]{tacchella_2022b}.  Over-plotted in Fig. \ref{fig:tform_zobs} are recent spectroscopic results of massive quiescent galaxies at $z > 1$ \citep[][]{kriek2016,carnall2022,hamadouche_connection_2023,antwi-danso_2025,kakimotor2024,Slob2024,nanayakkara_2025,degraaff2024}. The gray contour shows data from the large, ultra-deep, ground-based spectroscopic program VANDELS \citep[][for 114 massive quiescent galaxies at $1<z<1.3$]{hamadouche_connection_2023}. 
In the right-hand panel of Fig. \ref{fig:tform_zobs}, we show dashed lines of constant age at 0.5 Gyr, 1 Gyr and 2 Gyr. The \textit{DeepDive} galaxies lie along the 0.5 Gyr line, consistent with the ages obtained from \pipes\ and similar to recent results of massive high redshift quiescent galaxies  \cite{nanayakkara_2025,antwi-danso_2025,baker2925flamingojades}. We infer a similar stellar mass as \cite{carnall2024} for one of the galaxies observed as part of the EXCELS survey (ZF-UDS-6496, DD-80) and a marginally earlier formation time of $t_{\mathrm{form}}/\mathrm{Gyr}=0.99^{+0.03}_{-0.05}$, though still consistent to within 1$\sigma$. The majority of the \textit{DeepDive} galaxies formed the bulk of their stellar masses approximately 1 Gyr after the beginning of the Universe (from $t_{\mathrm{form}}$), reaching $\logM \sim 10.5-11$ by an average formation redshift of $z \sim 5$. Only one galaxy in our sample appears to have formed at $z\sim6$ (DD-96), and interestingly, this galaxy has also been quenched the longest; DD-96 quenched at $z \sim 4.9$ (corresponding to a quenching time of $t_{\mathrm{quench}} \simeq 1.21$). Overall, we find that quenching times are quite varied, though in good agreement across SFH parameterizations and recent literature \citep[][]{park2024b,baker2925flamingojades}. We do find that the galaxies in our {\it DeepDive} sample formed and quenched slightly later than those galaxies presented in  \cite{baker2925flamingojades}, though the sample on average is consistent with having long and short formation and quenching timescales. 

\subsubsection{Star-formation histories of the \textit{DeepDive} sample}

In Fig. \ref{fig:sfhs}, we present the star-formation histories and cumulative stellar-mass histories of our \textit{DeepDive} galaxies.  We find that most galaxies in our sample formed the bulk of their stellar mass rapidly , with typical formation timescales (calculated as the time between which the galaxy forms 20\% and 80\% of its stellar mass, $t_{80} - t_{20}$) between $\tau = 100 - 500$\,Myr (see Table \ref{tab:results_table2}). 
The right-hand panels for each galaxy shows the cumulative stellar mass history extrapolated down to redshift $z \sim 2$, with the SUSPENSE galaxies over-plotted \citep[][]{Slob2024}. The SUSPENSE sample, which consists of massive quiescent galaxies at $1<z<3$, provide an excellent comparison sample to the \textit{DeepDive} sample to trace the evolution of chemical abundances and star-formation histories (excluding any significant growth via mergers after quenching) from $z \sim 4$ to $z\sim 1$.  

\subsubsection{Comparisons between SFH parameterizations}\label{sec:sfh_comps}
In recent years, there has extensive discussion regarding the robustness of star-formation histories, stellar masses, and the timescales of formation and quenching inferred from SED fitting. 

We investigate the differences between four different SFH prescriptions in Fig. \ref{fig:sfh_comp_96} for all of the galaxies in our \textit{DeepDive} sample. Upon inspection, it appears that the majority of the SFH prescriptions agree well with each other within the 1-$\sigma$ uncertainty (shaded regions), with the exception of the \texttt{DenseBasis} \citep{iyer2019} SFH, which tends to be affected by the ``first-bin burstiness'' \citep[see][]{mcconachie_2025}. This does not appear to be a major concern for the other SFH models.
With the \texttt{DenseBasis} \citep{iyer2019} SFH model, we find that associated formation times are in good agreement with the other three SFH models, however, the timescales are offset lower by $\sim 30$\,Myr compared to our fiducial model values. When using equal Dirichlet parameters (which controls the smoothness of the smoothness between bins, $\alpha=3$) for the same number of bins, we find that the offset between formation times is on the order of $\sim150$\,Myr, resulting in older ages and longer formation timescales than the rest of the SFH models. 
This is because a significant fraction ($\sim$ 20 \%) of the galaxy's stellar mass is produced within the first half a Gyr of the Universe, thus extending the formation timescales to nearly the age of the Universe at observation.

We find that the stellar masses derived from each SFH is consistent within errors, and the fiducial continuity model from \pipes\ suggests formation $\sim 100$ Myr earlier on average than the bursty or parametric models, as demonstrated in the second and fourth rows of Fig. \ref{fig:sfh_comp_96}. This is in agreement with previous studies of MQGs at similar redshifts \citep{leja2019,baker2025b,mcconachie_2025}. 

We find that the \textit{DeepDive} galaxies do not demonstrate any significant early burstiness when comparing the bursty prior and our fiducial SFH models, though a few (DD-78, DD-196, DD-106, DD-115) appear to have undergone a strong starburst phase within the last 200 Myr prior to the epoch of observation. This may also suggest that our current measurements do not have the capacity to resolve very short timescale burstiness at earlier times. 
% the stellar populations of our galaxies are relatively older while the burstiness is better constrained by the combination of recent star formation and pre-existing older stars (traced by shallow Balmer breaks). 
Consistent with other studies, we find that the continuity model (both fiducial and bursty) tends to produce older ages than the parametric SFH models in \pipes\ \citep[see e.g.][]{leja2019,carnall2024,mcconachie_2025}. 
It can be seen in Fig. \ref{fig:sfh_comp_96} that though there is clear similarity between the formation and quenching times of each SFH, the double-power law model tends to fall off much more dramatically relative to the other three SFH prescriptions. This is likely because the parametrized double-power law model is biased towards producing most of the mass in a late, single burst of star formation. 
This is also observed in the distributions of stellar mass, metallicity and formation time presented in Fig. \ref{fig:corners} (for two randomly chosen galaxies in our sample); the double-power law SFH model tends to later formation times and slightly higher stellar metallicities compared to the non-parametric SFH models for all galaxies. 

Overall, it is reassuring to see that the non-parametric models have significant overlap in their posterior distributions for each galaxy. 
% In particular, the bursty and fiducial continuity SFH models agree very well for all galaxies for all derived properties (see Tables \ref{tab:results_table1} and \ref{tab:results_table2}), suggesting that these massive systems do not seem to demonstrate clear burstiness in their SFHs. 
% This could also potentially be attributed to the fact that we may not have the constraining power to determine the true burstiness of these galaxies;
It is interesting to note that the reduced $\chi^{2}$ values for all SFH models are the same within errors. Importantly, we find that the stellar masses, formation times and SFRs are robust against our different choice of model parameterizations.
We also find that across all SFH models, a significant number of our galaxies display high peak star-formation rates (SFR $>500 \ \mathrm{M_{\odot}}/\mathrm{yr}^{-1}$) before quenching, in line with the idea that these may be possible descendants of extremely dust-obscured star-forming galaxies observed at high redshift \citep{valentino2020}, consistent with number densities of DSFGs at $z>4$ \citep[e.g.][]{manning2025}. 

\subsubsection{Formation and quenching timescales from \pipes}\label{sec:timescales_pipes}
We report values for ages and star-formation timescales for each of the SFH model parameterizations in Table \ref{tab:results_table2}. Overall, we find good agreement for $t_{\mathrm{form}}$ between SFH models. A significant difference between all SFH prescriptions is the star-formation timescale. We define this as $\tau_{SF} = t_{80} - t_{20}$, the time it takes to increase the stellar mass from 20\% to 80\% of the total stellar mass formed. We find that, although the four SFH parameterizations perform well in reproducing similar formation times (see Table \ref{tab:results_table2}), $\tau_{\mathrm{SF}}$ varies widely; the double-power law SFH has the shortest formation timescales, with a median of $\tau = 67$\,Myr, and \texttt{DenseBasis} \citep{iyer2019} produces a longer median timescale of $\tau = 214$\,Myr, due to the early peak in star-formation for some of the galaxies. The fiducial SFH model has a median star-formation timescale of $\tau = 155$\,Myr, shorter than the \texttt{DenseBasis} SF timescales, but longer than the DPL and bursty SFH models.

\begin{figure*}
    \centering
\includegraphics[width=\linewidth]{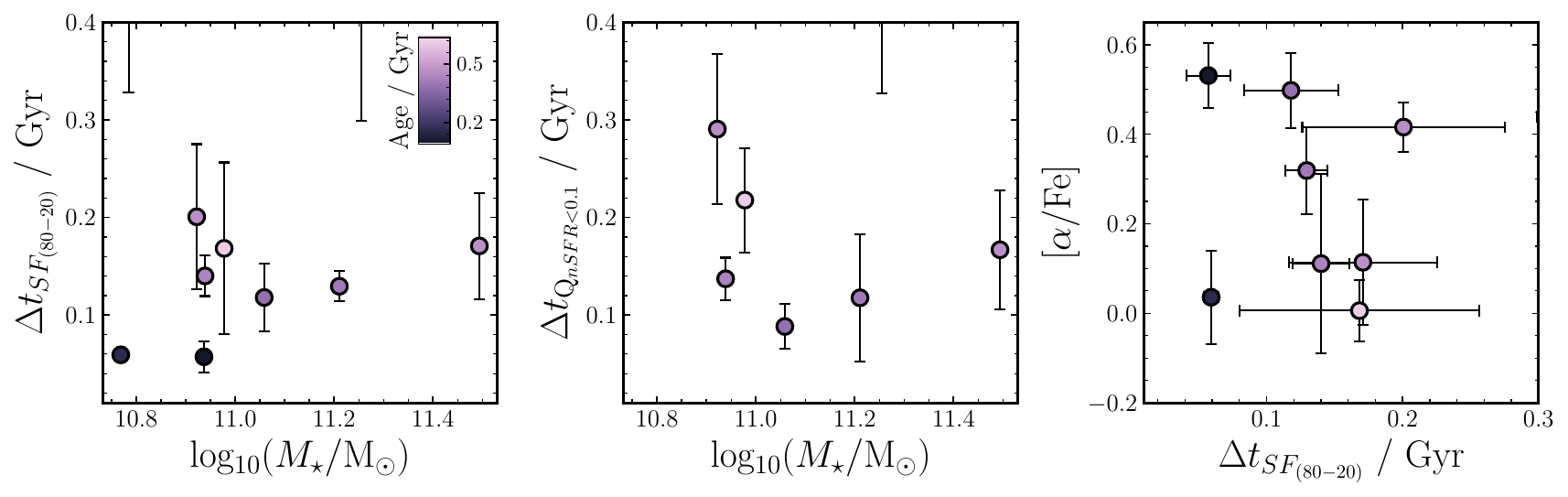}
    \caption{{\it Left and center:} Star-formation and quenching timescales versus stellar mass for our {\it DeepDive} sample from our fiducial \pipes\ fitting. The star-formation timescale is defined as the time it takes for a galaxy's stellar mass to rise from 20\% to 80\%, whereas the quenching timescale is the time between the formation ($t_\mathrm{form}$) and quenching ($t_{\mathrm{quench}}$) of a galaxy. $t_{\mathrm{quench}}$ is the time at which the galaxy's normalized SFR (nSFR) drops below 0.1. The two galaxies DD-78 and DD-111 do not have quenching timescales, most likely due to the effect of the AGN on their SFHs. {\it Right:} [$\alpha$/Fe] versus formation timescale. No strong correlation is visible between formation timescale and $\alpha$-enhancement in our {\it DeepDive} sample, demonstrating the varied formation and quenching pathways of massive quiescent galaxies at $z>3$.}
    \label{fig:timescales_bp}
\end{figure*}

We additionally consider quenching timescales for all galaxies in the {\it DeepDive} sample. The quenching time is defined as the age of the Universe at which the galaxy's normalized SFR falls below 0.1 derived from \pipes. The quenching timescale is therefore the difference between the quenching time and the formation time. We show our derived quenching times and timescales along with the formation times and timescales for our fiducial \pipes\ results in the left and center panels of Fig. \ref{fig:timescales_bp}. 

The quenching timescales derived for our sample similarly reflect the diverse formation and quenching pathways -- the {\it DeepDive} galaxies quenched on timescales of $<300$\,Myr with a median quenching timescale of $\simeq252$\,Myr for our fiducial run (bursty $\simeq 214$\,Myr, \texttt{DenseBasis} $\simeq 270$\,Myr, double-power law $\simeq 278$\,Myr). The quenching timescales derived for all SFH parameterizations are therefore in good agreement, with the bursty SFH producing somewhat shorter quenching timescales than the rest of the parameterizations. Finally, when considering the correlation between $\alpha$-enhancement and star-formation timescale, there is a tentative trend towards galaxies with higher [$\alpha$/Fe] having shorter star-formation timescales, though we find this is not statistically significant ($p-\mathrm{value}=0.96$), due to the low numbers and narrow redshift range probed by our sample. Similarly, we find no statistically significant correlation in our sample between [$\alpha$/Fe] and quenching timescale ($p-\mathrm{value}=0.56$).  

Reassuringly, we find that our results for formation and quenching timescales agree well with recent works at similar early times. \cite{baker2925flamingojades} and \cite{park2024b} show that massive quiescent galaxies at $z>2$ have both long and short SFHs; galaxies that formed and quenched early and rapidly, those that formed and quenched on more extended timescales, as well as those that formed in very short starbursts \citep[as in ][]{carnall2024}. In our sample, we see evidence of galaxies with extended formation timescales (e.g., DD-170, DD-80) as well as galaxies that formed extremely fast in rapid bursts (e.g., DD-78, DD-115). The observed rapid timescales may indicate quenching via AGN feedback, in line with the duty cycles measured for AGN activity (see Section \ref{sec:SFtimescalesvalpha}).

\begin{deluxetable*}{ccccccccc}
\tablecaption{Formation times ($t_{\mathrm{form}}$, age of the Universe at which 50\% of a galaxy's stellar mass is formed) and star-formation timescales ($\tau_{SF} = t_{80} - t_{20}$, where $t_{20}$ [$t_{80}$] is the  age of the Universe at which 20\% [80\%] of a galaxy's stellar mass formed) for the different SFH prescriptions used in \pipes\ described in Section \ref{sec:sfh_comps}. Uncertainties given are the 16th and 84th percentiles on the posterior. 
}
\label{tab:results_table2}
\tablecolumns{9}
\tablehead{
\colhead{} & 
\multicolumn{2}{c}{\texttt{DenseBasis}}& 
\multicolumn{2}{c}{Double-power Law} & 
\multicolumn{2}{c}{Bursty} & \multicolumn{2}{c}{Fiducial} \\
\colhead{DD-ID} & \colhead{$t_{\mathrm{form}}$/Gyr}  & \colhead{$\tau_{\mathrm{SF}}$/Myr} & \colhead{$t_{\mathrm{form}}$/Gyr}  & \colhead{$\tau_{\mathrm{SF}}$/Myr} & \colhead{$t_{\mathrm{form}}$/Gyr}  & \colhead{$\tau_{\mathrm{SF}}$/Gyr} & \colhead{$t_{\mathrm{form}}$/Gyr}  & \colhead{$\tau_{\mathrm{SF}}$/Myr} }
\startdata
78 & $1.27^{+0.03}_{-0.02}$ & 335$^{+45}_{-49}$ & $1.33^{+0.01}_{-0.02}$ & $51.3^{+7.0}_{-4.8}$  & $1.42^{+0.01}_{-0.01}$ & $15.4^{+1.4}_{-0.9}$  & $1.41^{+0.01}_{-0.02}$ & $57.2^{+13.1}_{-19.1}$  \\
80 & $1.06^{+0.06}_{-0.08}$ & $225^{+470}_{-55}$ & $1.13^{+0.03}_{-0.03}$& $85.3^{+54.4}_{-11.6}$ & $1.11^{+0.02}_{-0.02}$ & $113^{+50}_{-33}$ & $1.08^{+0.04}_{-0.04}$ & $200^{+88}_{-61}$\\
96 & 1.10$^{+0.06}_{-0.07}$ & $203^{+17}_{-50}$ & $1.06^{+0.03}_{-0.03}$ & $20.0^{+44.2}_{-6.3}$ & $1.01^{+0.01}_{-0.01}$ & $104^{+13.7}_{-3.7}$  & $1.00^{+0.04}_{-0.07}$ & $168^{+121}_{-55}$\\
106 & $1.18^{+0.03}_{-0.04}$ & $178^{+35}_{-50}$ & $1.20^{+0.03}_{-0.06}$ & $84.3^{+10.1}_{-6.2}$ & $1.17^{+0.03}_{-0.01}$ & $81.4^{+39.0}_{-5.8}$ & $1.21^{+0.01}_{-0.03}$ & $130^{+8.0}_{-22.0}$ \\
111 & $1.25^{+0.03}_{-0.04}$ & $353^{+58}_{-49}$  & $1.26^{+0.07}_{-0.17}$ & $241^{+326}_{-148}$  & $1.32^{+0.04}_{-0.07}$  & $117^{+87}_{-48}$   & $1.09^{+0.09}_{-0.08}$ & $556^{+251}_{-204}$  \\
115 & $1.42^{+0.02}_{-0.06}$  & $233^{+50}_{-53}$  & $1.47^{+0.00}_{-0.01}$ & $36.2^{+4.8}_{-2.6}$ & $1.46^{+0.01}_{-0.01}$ & $43.9^{+15.3}_{-8.2}$ & $1.46^{+0.01}_{-0.01}$  & $59.3^{+2.8}_{-7.8}$ \\
134 & $1.27^{+0.02}_{-0.02}$ & $153^{+44}_{-50}$  & $1.28^{+0.04}_{-0.04}$ & $25.1^{+24.0}_{-17.1}$ & $1.32^{+0.01}_{-0.02}$ & $64.8^{+36.6}_{-7.8}$ & $1.29^{+0.03}_{-0.02}$ & $118^{+17}_{-52}$  \\
170 & $1.30^{+0.05}_{-0.05}$ & $264^{+46}_{-30}$  & $1.24^{+0.06}_{-0.08}$ & $425^{+110}_{-101}$ & $1.27^{+0.05}_{-0.06}$ & $290^{+90}_{-96}$  & $1.17^{+0.07}_{-0.11}$ & $467^{+186}_{-150}$ \\
179 & $1.37^{+0.03}_{-0.08}$ & $203^{+47}_{-64}$ & $1.34^{+0.07}_{-0.04}$ & $81.7^{+67.3}_{-11.1}$ & $1.35^{+0.01}_{-0.02}$ & $88.6^{+17.2}_{-4.2}$  & $1.36^{+0.04}_{-0.08}$ & $170^{+79}_{-30}$  \\
196 & $1.36^{+0.03}_{-0.03}$  & $204^{+38}_{-66}$ & $1.41^{+0.02}_{-0.02}$ & $17.3^{+23.6}_{-10.0}$ & $1.37^{+0.02}_{-0.01}$ & $88.6^{+32.4}_{-6.2}$ & $1.40^{+0.02}_{-0.03}$ & $140^{+7}_{-34}$  \\
\enddata
\end{deluxetable*}

\subsection{Metallicities of the \textit{DeepDive} sample}\label{sec:abundances}

In general, studies have shown that massive quiescent galaxies demonstrate evidence for enhanced $\alpha$-element abundances (e.g, O, Mg, Si, Ca, S, Na, Ti); at higher redshifts, these abundance patterns are tightly correlated with the star-formation history of a galaxy, due to the timescales on which $\alpha$ elements are produced in CCSNe relative to iron or iron-peak elements (from Type Ia SNe), which, in turn, are also dependent on the initial mass function.  To obtain independent empirical estimates on the chemical enrichment history, and therefore formation and quenching histories of the \textit{DeepDive} galaxies, we measure metallicity and $\alpha$-enhancement using the Pythonic element abundance fitting code, \alfa\footnote{https://github.com/alizabeverage/alfalpha} \citep[for more details, see][]{beverage2025}. This code is based on the \texttt{fortran} Absorption Line Fitter \citep[see][]{conroy_2018} with the MILES+IRTF+MIST SSP models, suitable for measuring elemental abundances for older ($>1$ Gyr) stellar populations. Importantly, these models also use theoretical response functions to vary individual elemental abundances.

Given the young galaxies analyzed in this paper, \alfa\ also has functionality to enable extending the modeling to younger stellar populations with the semi-empirical MILES models \citep[sMILES, see][]{knowles2023}.  These models make use of the empirical MILES stellar library to provide spectra with varying $\alpha$-enhancements over a large range of stellar ages, metallicities and [$\alpha$/Fe] computed for five initial mass functions.  That said, these models do not include individual theoretical response functions as in the \cite{conroy_2018} models, so we are only able to measure total [$\alpha$/Fe] values rather than the commonly used [Mg/Fe] value.

The appeal of using \alfa\ with sMILES lies in the ability to model abundance measurements for SSPs down to 30 Myr, currently unavailable for the \cite{conroy_2018} models and response functions. However, as \alfa\ only assumes a single burst of star-formation, the stellar population age is also the time at which the burst occurred. Future work implementing improved stellar population models -- such as the soon-to-be publicly-available $\alpha$-MC models \citep[][]{park2024a} -- in SED fitting codes will provide more sophisticated modeling for a range of star-formation histories. This will enable us to finally break the degeneracy between ages, metallicities and star-formation histories when fitting galaxies with non-solar abundance patterns, allowing for more reliable and less biased results from SED fitting.

As the sMILES models only return total solar-scaled metallicities ([Z/H]), we first must convert to non-solar abundances (to retrieve [Fe/H]) using Eq. \ref{eqn:fehconv} below, (see Eq. 6 in \citealt{knowles2023}): 

\begin{equation}\label{eqn:fehconv}
    \mathrm{[Fe/H]} = [Z/\mathrm{H}] - a[\alpha/\mathrm{Fe}] - b[\alpha/\mathrm{Fe}]^{2},
\end{equation}
where the coefficients $a = 0.66154 $ and $b = 0.20465$. 
The values we obtain for [$\alpha/\mathrm{Fe}$], [Fe/H], age  and velocity dispersion ($\sigma / \mathrm{km\ s^{-1}}$) derived from \alfa\, are presented in Table \ref{tab:results_table}. The sMILES models provide a rest-frame wavelength coverage of $3540.5 - 7409.6$ \AA, however, we fit our individual spectra within the region $3700 - 6400$ \AA\ for consistency with \cite{beverage2025}. Although the models are computed for five initial mass function types, for consistency with the literature and our \texttt{Bagpipes} results, we default to a universal \cite{kroupaimf} IMF. Similar to our \texttt{Bagpipes} fitting procedure, we also mask AGN emission lines and the Na \textsc{I} doublet feature at $\sim$5895 \AA,  due to the additional ISM dependence of the latter. 

\begin{deluxetable*}{ccccccc}
\tablecaption{Derived properties from the \texttt{alf}$\alpha$ (using the sMILES models, see \citealt{knowles2023}) fits for the \textit{DeepDive} galaxies, in order of increasing spectroscopic redshift. Galaxies 78 and 111 are not shown due to significant AGN emission features in their spectra affecting the \alfa\ fitting. The third column displays the ages of each galaxy derived from \alfa, not the formation times (age of Universe - age from \alfa). Uncertainties given are the 16th and 84th percentiles on the posterior.
}
\label{tab:results_table}
\tablecolumns{7}
\tablehead{
\colhead{DD-ID} & 
\colhead{$z_{\mathrm{spec}}$} & 
\colhead{Age / Gyr}  & 
\colhead{[$\alpha$/Fe]} & 
\colhead{[Fe/H]} & \colhead{[Z/H]} &
\colhead{$\sigma$/km s$^{-1}$}}
\decimals
\startdata
196 & 3.4332 & $0.36^{+0.06}_{-0.05}$ & $0.11_{-0.19}^{+0.23}$ & $0.04_{-0.21}^{+0.25}$ & $0.15_{-0.18}^{+0.08}$ & $273.20_{+23.19}^{+23.79}$\\
179 & 3.4472 & $0.49_{-0.05}^{+0.06}$ & $0.11_{-0.13}^{+0.15}$ & $0.05_{-0.13}^{+0.10}$ & $0.15_{-0.13}^{+0.08}$ & $319.67_{-24.90}^{+26.92}$ \\
170  & 3.4894 & $0.51_{-0.06}^{+0.10}$ & $0.44_{-0.23}^{+0.11}$ & $-0.49_{-0.18}^{+0.23}$ & $-0.18_{-0.13}^{+0.21}$ & $382.58_{-27.83}^{+28.51}$  \\
134  & 3.7130 & $0.39_{-0.03}^{+0.03}$ & $0.50_{-0.10}^{+0.07}$  & $-0.20_{-0.07}^{+0.08}$ & $0.18_{-0.06}^{+0.06}$ & $271.80_{-15.57}^{+16.43}$  \\
115  & 3.7766 & $0.15_{-0.01}^{+0.02}$ & $0.04_{-0.10}^{+0.11}$ & $0.10_{-0.09}^{+0.08}$ & $0.13_{-0.08}^{+0.07}$ & $289.53_{-8.79}^{+8.71}$  \\
106 & 3.8307 & $0.43_{-0.04}^{+0.06}$ & $0.32_{-0.10}^{+0.09}$ & 
$-0.07_{-0.09}^{+0.06}$ & $0.17_{-0.10}^{+0.07}$ & $332.05_{-18.16}^{+18.44}$ \\
96  & 3.8676 & $0.65_{-0.07}^{+0.06}$ & $0.01_{-0.07}^{+0.07}$ & 
$-0.09_{-0.09}^{+0.11}$ & $-0.09_{-0.10}^{+0.14}$ & $359.45_{-15.68}^{+15.66}$\\
80 & 3.9879 & $0.44_{-0.02}^{+0.03}$ & $0.42_{-0.05}^{+0.06}$ &
$-0.10_{-0.07}^{+0.05}$ & $0.22_{-0.05}^{+0.03}$ & $406.46_{-13.65}^{+14.18}$\\
\enddata
\end{deluxetable*}

\begin{figure*}
\centering
\includegraphics[width=\linewidth]{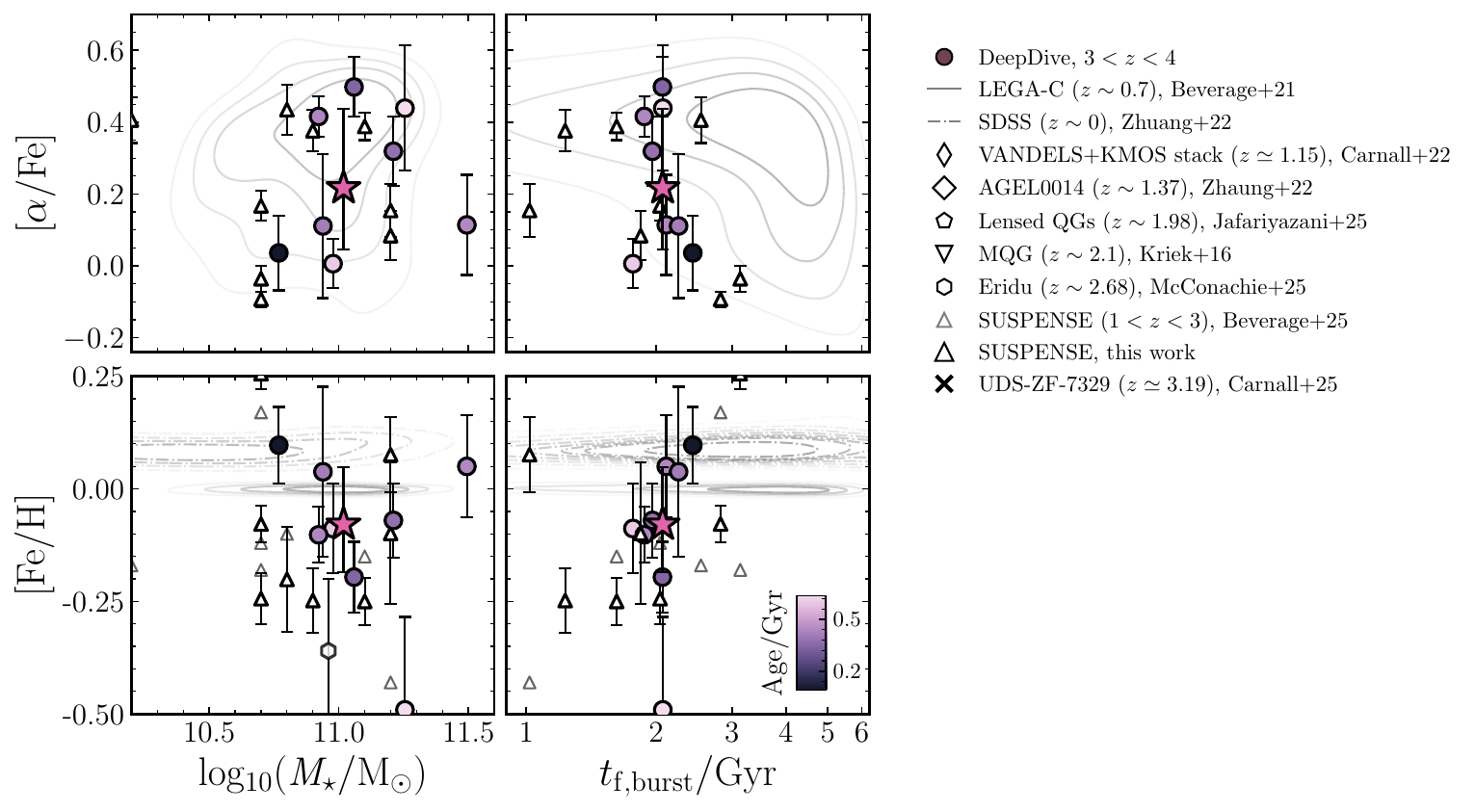}
    \caption{[$\alpha$/Fe] and [Fe/H] versus stellar mass and formation time for the \textit{DeepDive} galaxies colored by mass-weighted age from \pipes, excluding the two galaxies with strong AGN emission lines (see Section \ref{agnemission}). The pink star is the median value (of the points) for the {\it DeepDive} galaxies. For consistency, the formation time shown is the difference between the age of the Universe at the observed galaxy redshift and the age of the galaxy from \alfa. Triangles with the solid outline represent the SUSPENSE [$\alpha$/Fe] and [Fe/H] values measured in this work using \alfa. We also over-plot the LEGA-C DR3 [$\alpha$/Fe] measurements (solid-line gray contour). The [$\alpha$/Fe] values for the \textit{DeepDive} sample shows strong diversity, though in good agreement with lower redshift studies (i.e. with results from the LEGA-C sample at $z \sim 0.7$). We find overall lower [Fe/H] values on average for our sample than $z\sim 0.7$ and $z\sim 0$ studies, consistent with \cite{beverage2025}. Mass uncertainties are not shown for clarity (see Table \ref{tab:results_table1}).
}
    \label{fig:afe_masstform}
\end{figure*}

\subsubsection{$\alpha$-enhancement in massive galaxies at $z>3$}
In Fig. \ref{fig:afe_masstform}, we present our [$\alpha$/Fe] and [Fe/H] measurements versus stellar mass and formation time ($t_{\mathrm{form}}$) for the {\it DeepDive} sample, alongside several literature measurements for massive quiescent galaxies. Our results demonstrate that MQGs at high redshift are $\alpha$-enhanced, as expected given the shorter formation and quenching timescales at early times. 
On average, the \textit{DeepDive} galaxies display a range of [$\alpha$/Fe] values (see Table \ref{tab:results_table}), with a median \hbox{[$\alpha$/Fe] = $0.22^{+0.22}_{-0.17}$}. [$\alpha$/Fe] values obtained from our fitting are lower than those found in current literature studies using the same approach \citep{kriek_2019} and code \citep[using \alfa,][]{beverage2025,mcconachie_2025}, with different models. However, we emphasize that we are measuring \textit{total} $\alpha$-enhancement rather than the [Mg/Fe] abundance ratio that is commonly quoted. In massive quiescent galaxies at high redshifts, Ca and Si are typically (less than) solar values, meaning that [Mg/Fe] will always be higher than the combined average of all $\alpha$-elements \citep[see e.g.][]{beverage2025}. We discuss this in more depth in Section \ref{sec:discussion}. 

To provide a self-consistent comparison of the $\alpha$-enhancements of massive quiescent galaxies, we fit the SUSPENSE spectra (reduced by the SUSPENSE team, \citealt{Slob2024}) by applying the same method to the SUSPENSE galaxies with \alfa\, as used for our \textit{DeepDive} analysis.  

In the bottom panels of Fig. \ref{fig:afe_masstform}, we show the [Fe/H] values derived for the {\it DeepDive} galaxies, finding mostly sub-solar [Fe/H] values, with a median \hbox{[Fe/H] = $-0.08^{+0.13}_{-0.11}$}, in agreement with single object determinations at $ z \simeq 3.2$ \citep[][]{carnall2024} and results at $1<z<3$ from \cite{beverage2025}. This value is lower (by $\sim 0.1$ dex) than [Fe/H] values for studies at lower redshift (e.g., LEGA-C, $\sim 0.2$ dex lower than SDSS), though consistent with our understanding that galaxies at higher redshift have lower metallicities than their low-redshift counterparts at fixed stellar mass \citep[][]{beverage2024}. We obtain a lower median value for $\alpha$-enhancement than those reported by recent studies of massive quiescent galaxies at high redshift, though still within $1\sigma$ \citep[e.g.][for $\mathrm{[Mg/Fe]} = 0.42^{+0.19}_{-0.17}$ at $z \simeq 3.2$]{carnall2024}, and not at odds with other findings considering that we measure the average enhancement for all $\alpha$ elements, instead of [Mg/Fe]. For example,
\cite{beverage2025} find $\alpha$-enhancement values of \hbox{[Mg/Fe] = $\sim 0.4-0.5$} at $1<z<3$ and \cite{mcconachie_2025} report a value of [Mg/Fe] = $\sim 0.65$ for a single massive quiescent galaxy at $z \sim 2.5$.

Unfortunately, the sMILES models do not incorporate element response functions and so it is not possible to vary individual metal abundances as performed in other studies \citep{conroy_2018,kriek_2019,carnall2023,carnall2024,2jafariyazani_2025}. However, these results, combined with results from the literature, suggest that our overall conclusion that galaxies at $z>3$ are $\alpha$-enhanced is robust, even though the absolute values may be different due to the choice of models and/or definitions (including the different solar-abundance scales used between different models). We discuss this in more detail in Section \ref{sec:discussion}.

When considering the \textit{DeepDive} sample alone, we find no strong statistical evidence for correlations between stellar mass and abundances or formation time. In Fig. \ref{fig:afe_masstform}, our galaxies appear to populate a narrow stellar-mass range of $< 1$ dex. However, when placed in the context of lower-redshift results, we can see that our \textit{DeepDive} galaxies are consistent with having higher [$\alpha$/Fe] values due to having formed at earlier times than MQGs at lower stellar masses with later formation times. Overall, this suggests evidence for more massive galaxies having higher values for $\alpha$-enhancement, potentially an effect of downsizing (see \citealt{cowie1996downsizing,fontanot2017,gallazzi2025}), implying downsizing may already be in place at these early times. We observe a negative trend with formation time (t$_{\mathrm{f, burst}}$ = age of the Universe -- age from \alfa), such that galaxies that formed earlier also have higher $\alpha$-enhancements (when comparing results at all redshifts). Interestingly, this negative trend holds with the formation times obtained from our \pipes\ fitting; galaxies with lower [$\alpha$/Fe] formed more slowly, and thus later, in the Universe than those with higher values, due to the long, delayed timescales on which Type Ia SNe enrich the ISM with Fe (see Section \ref{sec:SFtimescalesvalpha}). These trends are also visible in the SUSPENSE sample; the similarities between the distributions of both datasets are clear and additionally suggest that there is little evolution between $z \sim 4$ to $z\sim 1$. 
We additionally compare our results to the [$\alpha$/Fe] values measured for the LEGA-C DR3 galaxies \citep[using the \citealt{conroy_2018} models, ][]{beverage2023}, shown as the gray contour in Fig. \ref{fig:afe_masstform}. We find that our {\it DeepDive} measurements are in good agreement with the LEGA-C and SUSPENSE galaxies, with strong overlap between all three samples. To quantify the overlapping distributions, we perform a 2-sample Kolmogorov-Smirnov (KS) test to investigate whether the {\it DeepDive} [$\alpha$/Fe] values are drawn from the same distributions as LEGA-C and SUSPENSE. We find  $p-\mathrm{value} = 0.69$ and  $p-\mathrm{value} = 0.86$ for the two-sample KS tests between the LEGA-C and SUSPENSE samples, respectively, allowing us to confirm the null hypothesis that the {\it DeepDive} galaxies are drawn from the same distributions as the the $0.6<z < 0.8$ and $1<z<3$ samples. This thus implies weak evolution in [$\alpha$/Fe] between $z \sim 4$ to $z\sim0.6$.

\subsection{The stellar mass-metallicity relation}
\begin{figure}
\centering
\includegraphics[width=\linewidth]{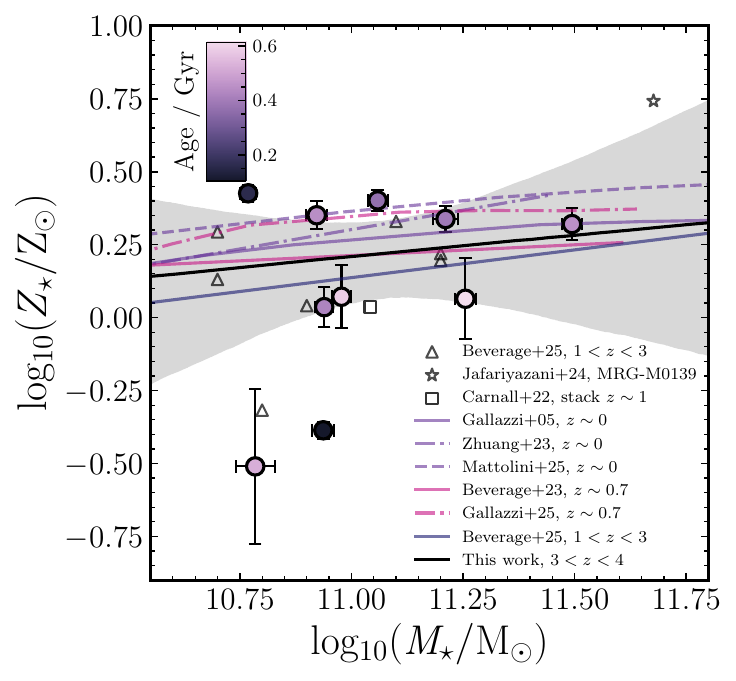}
    \caption{Stellar mass versus metallicity from \texttt{Bagpipes}, colored by mass-weighted age for all quiescent galaxies in our sample. Our best-fit relation is in good agreement with recent, lower redshift results, albeit with a slightly lower normalization, though still within 1$\sigma$, implying little evolution in the MZR between $z\sim 4$ to $z\sim 0$ for massive quiescent galaxies.}
    \label{fig:massmet}
\end{figure}

Using the total stellar metallicities ($\mathrm{log}_{10}(Z_{\star}/Z_{\odot})$) obtained from our \pipes\ fitting, we measure the mass-metallicity relation (MZR) for the ten \textit{DeepDive} galaxies in our sample. While we use the \pipes\ results for consistency when providing comparisons with the literature, we point out that our \alfa\ results for [Z/H] are in excellent agreement with \pipes\ (within $1\sigma$). We show our \pipes\ results in Fig. \ref{fig:massmet} along with our best-fit relation (solid black line). The shaded gray area corresponds to the 1$\sigma$ uncertainty on the derived relation.  We over-plot individual galaxy results from recent studies of massive quiescent galaxies from $z\sim 0$ to $z\sim 3$.  The triangles correspond to values from SUSPENSE \citep{beverage2025} after converting their [Fe/H] values to total metallicities using the relation from \cite{thomas2003}:

\begin{equation}
    \mathrm{log}_{10}(Z_{\star}/Z_{\odot}) = 0.94 \mathrm{[Mg/Fe]} + \mathrm{[Fe/H]}.
\end{equation}

We also show the $z\sim 0$ relations  \citep{gallazzi2005,zhuang2023,mattolini2025}, $z\sim 0.7$ (\citealt{beverage2023},\citealt{gallazzi2025}) and $1<z<3$ relations \citep{beverage2025}, respectively. 

We derive a best-fitting relation between stellar mass and total stellar metallicity (from \pipes) for all {\it DeepDive} galaxies by fitting a linear relationship using the \texttt{dynesty} \citep[][]{speagle2020,koposov2015} nested sampling \citep[][]{skilling2004,skilling2006} package (employing the dynamic nested sampling algorithm for improved parameter estimation, \citealt{higson2019}). Our best-fitting MZR is

\begin{equation}
    \mathrm{log}_{10}(Z_{\star}/Z_{\odot}) = 0.15^{+0.46}_{-0.40}  \mathrm{log}_{10}\left(\frac{M_{\star}}{10^{11} M_{\odot}}\right) + 0.21^{+0.08}_{-0.12}, 
\end{equation}

where the errors on the slope and offset are $1\sigma$ uncertainties.
Our best-fit relation is flat, in good agreement with the MZR at $z\sim 1 - 3$ discussed in \cite{beverage2025}, with our results implying very little evolution in the MZR from $z \sim 4$ to $z \sim 0$ \citep[see also][for light-weighted MZR evolution]{mattolini2025,gallazzi2025}. The normalization of our MZR is lower than $z \sim 0$ at studies by $\simeq0.10$ dex (at a fixed stellar mass of %$\sim0.11^{0.08}_{0.10}$ dex-
$\logM=11.00$, compared to the $z\sim 0$ relations from \citealt[][]{gallazzi2005,zhuang2023}). However, this offset between the $z\sim 4$ and $z \sim 0$ relations is only at the 1$\sigma$ level, suggesting little evidence for strong evolution at these high stellar masses and redshifts. 

The stellar metallicities obtained for the \textit{DeepDive} galaxies are in agreement with values found for massive quiescent galaxies at $z>3$ and in agreement with the values found for the SUSPENSE galaxies from \cite{beverage2025}. Although there is large uncertainty on our best-fit relation, this is understandable given the smaller size and narrow stellar-mass range of our sample compared to lower redshift studies. Larger combined archival samples from multiple programs, such as those presented in \cite{Ito_2025} will allow us to provide improved measurements of the MZR in the future. However, as a whole, our results place important constraints on the mass-metallicity evolution of high-redshift quiescent galaxies. 

\section{Discussion} \label{sec:discussion}
In this section, we discuss the results of our \pipes\ and \alfa\ fitting to the \textit{DeepDive} galaxies. Specifically, we explore  timescales and potential quenching mechanisms for the {\it DeepDive} sample. We also discuss the caveats associated with the models used in this work.

\subsection{Star-formation timescales inferred from $\alpha$-enhancements}

In the previous section, we showed that the \textit{DeepDive} sample spans a diverse range of formation and quenching timescales, independent of the SFH parametrization chosen, within 1$\sigma$. This therefore has strong implications on the various formation and quenching mechanisms that galaxies at $z \sim 3 - 4$ undergo. In addition, the chemical abundance patterns of galaxies at high redshift have strong implications on the evolution of galaxies; $\alpha$-enhancement correlates with star-formation timescale, allowing us to constrain the onset and cessation of their star formation. 

In Fig. \ref{fig:afe_masstform} we present the distribution of our \textit{DeepDive} sample for [$\alpha$/Fe] and [Fe/H] versus stellar mass and formation time (the age of the Universe at the galaxy's observed spectroscopic redshift minus the burst age obtained from \alfa). Although in general, higher redshift galaxies are more massive and appear to have on average higher [$\alpha$/Fe] than lower redshift samples \citep[][]{fontanot2017,gallazzi2025}, we find no evidence of this when considering the \textit{DeepDive} sample \textit{alone}, most likely due to the narrow range of ages and stellar masses probed by our sample at these high redshifts. However, we do find that the median [Fe/H] for our sample is almost $\sim 0.1$ dex lower than [Fe/H] from LEGA-C (at $z \sim 0.7$) and $\gtrsim0.2$-dex lower than $z\sim 0$ results (from SDSS), suggesting there is some evolution in the [Fe/H]. 
Our results are not inconsistent with results from analyses at lower redshifts; although we find that, overall, the majority of the \textit{DeepDive} galaxies are $\alpha$-enhanced (see also Table \ref{tab:results_table}), the range of [$\alpha$/Fe] values for the sample is large. 
The large scatter in [$\alpha$/Fe] for the \textit{DeepDive} and SUSPENSE results presented in this paper may be partly explained by the expected spread due to [$\alpha$/Fe] being an average over all $\alpha$ elements. \cite{beverage2025} demonstrates that not all $\alpha$ elements have the same abundances, possibly contributing to the observed scatter in both the \textit{DeepDive} and SUSPENSE samples.   
 
Moreover, this scatter in [$\alpha$/Fe] may be an effect of the varying star-formation histories exhibited by these galaxies. At lower redshifts ($z\sim1-3$), massive quiescent galaxies are found to have a range of star-formation and quenching timescales \citep[][]{carnall2019,hamadouche_connection_2023,Slob2024}, and indeed, the galaxies at $z>3$ presented in this paper show a range of star-formation timescales and quenching timescales (see Section \ref{sec:timescales_pipes} and Fig. \ref{fig:timescales_bp}). The broad range of [$\alpha$/Fe] values found for the \textit{DeepDive} sample may therefore reflect the multiple formation and quenching mechanisms and timescales apparent at high redshifts.
Similarly, the [O/Fe] of simulated massive quenched galaxies spans a range of $\sim0.25$ dex \citep[e.g. GAEA,][]{delucia2025}, with a median $\mathrm{[O/Fe]}=0.33$. Interestingly, massive quenched galaxies in Magneticum \citep[][]{kimmig2025} have lower average $\alpha$-enhancement with a mean of $\mathrm{[O/Fe]}=0.14$. This diversity may suggest either a range of formation and quenching timescales for galaxies at $3<z<4$, similar to lower redshift studies \citep[e.g.][]{carnall2022,hamadouche_connection_2023}, indicative of multiple formation and quenching pathways. For the SUSPENSE sample, we find a median [$\alpha$/Fe] = $0.17^{+0.23}_{-0.17}$, marginally lower than the \textit{DeepDive} sample, though still within 1$\sigma$. Overall, our results imply weak evolution between $z \sim 4$ to $z \sim 1$. 

To visualize the star-formation timescales ($\Delta t$) predicted for our sample from [$\alpha$/Fe], we use the equation from \cite{thomas2005}. This assumes a closed-box model, where selective mass loss does not occur and thus does not affect the [$\alpha$/Fe] ratio. Additionally, the SFHs are assumed to be parametrized by simple Gaussians.
\begin{equation}\label{eq:afe_equation}
    [\alpha/\mathrm{Fe}] = \frac{1}{5} - \frac{1}{6}\mathrm{log}(\Delta t / \mathrm{Gyr}).
\end{equation}
Following \cite{beverage2023}, the formation time calculated from \alfa\ is set as the mean of the distribution and the FWHM corresponds to the star-formation timescale, $\Delta t$. 
\begin{figure}
    \centering
    \includegraphics[width=\linewidth]{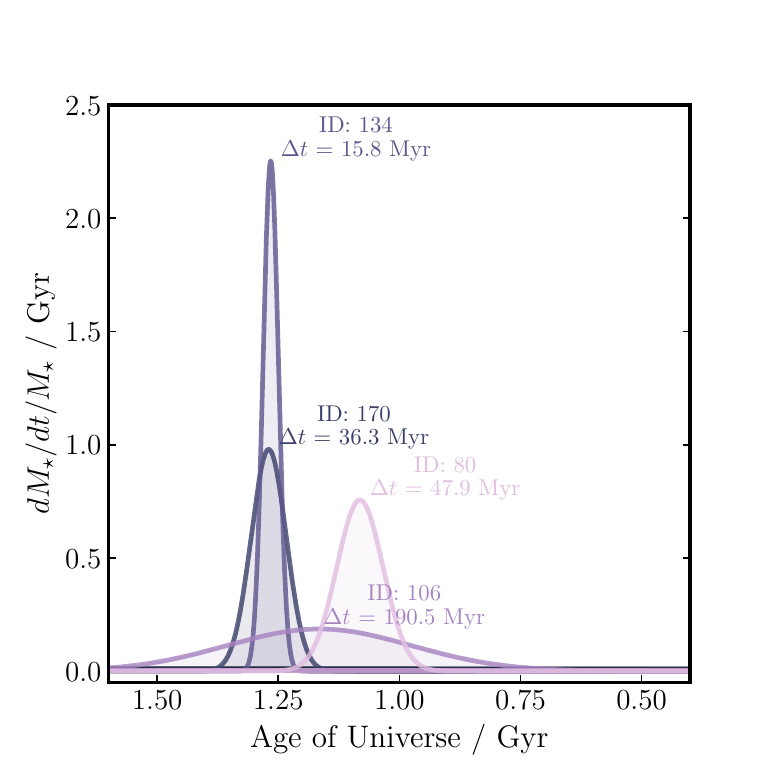}
    \caption{The star-formation timescales predicted for all 8 objects with [$\alpha$/Fe] values, calculated from Eq. \ref{eq:afe_equation}. The FWHM of the Gaussian distribution corresponds to the SF timescale, $\Delta t$, centered on each galaxy's formation time. The SF timescales based on our [$\alpha$/Fe] measurements vary widely, with 4 galaxies having timescales predicted to be $\gtrsim 4$ Gyr (distributions that appear flat in the figure). The long timescales predicted for some galaxies arise from the fact that the [$\alpha$/Fe] values are low in comparison to expected $\alpha$-enhancement values for massive high-redshift quiescent galaxies.}
    \label{fig:afetimescles}
\end{figure}

In Fig. \ref{fig:afetimescles}, we present the inferred star-formation histories based on our [$\alpha$/Fe] values. In general, the timescales predicted from our chemical abundances are shorter for galaxies with higher [$\alpha$/Fe], and those with lower [$\alpha$/Fe] values are much longer than the time elapsed between formation and observation.  
However, the possibility of longer timescales such as those at low $z$ is ruled out due to the short timescales on which these galaxies must have formed at these high redshifts. 
For example, DD-115 which has [$\alpha$/Fe]$=0.04^{+0.09}_{-0.10}$ has a SF timescale of $9.12$\,Gyr (compared with a timescale from \pipes\ of $\tau_{\mathrm{SF}} \simeq 59$\,Myr). For the galaxies with [$\alpha$/Fe]$>0.2$, we find that the SF timescales derived from our [$\alpha/$Fe] values are also much shorter than those derived from \pipes\, consistent with findings in \cite{beverage2025}. On average, the SF timescale derived for the median [$\alpha$/Fe] ($\simeq0.22$) is $\tau \simeq 750$\,Myr, longer than the median timescale determined with \pipes\ ($\tau \simeq 170$\,Myr).
However, it is important to note that this is an oversimplified model, and does not reflect the complex star-formation histories and evolutionary pathways (e.g., previous bursts of star-formation) seen at early times. 

The timescales inferred from our [$\alpha$/Fe] measurements are short, and much shorter than the timescales obtained from \pipes. Our results suggest that [$\alpha$/Fe] alone cannot be used as an indicator of star-formation timescales within massive quiescent galaxies at high redshift; the influence of a variable IMF \citep[e.g.][]{tinsley1979}, and merger/accretion events may contribute greatly to the observed trends \citep[][]{fontanot2017,delucia2017}. Specifically, a variable IMF can change the ratio of CCSNe to SNIa in distant quiescent galaxies which can affect the abundances of $\alpha$ to iron-peak elements produced \citep[][]{fontanot2017}. Additionally, the inclusion of different feedback modes in theoretical models is needed to understand in more detail the diversity of physical properties of early massive quiescent galaxies. Future work will explore more complex formation scenarios to understand their detailed chemical evolution. 

\begin{figure*}
    \centering
    \includegraphics[width=\linewidth]{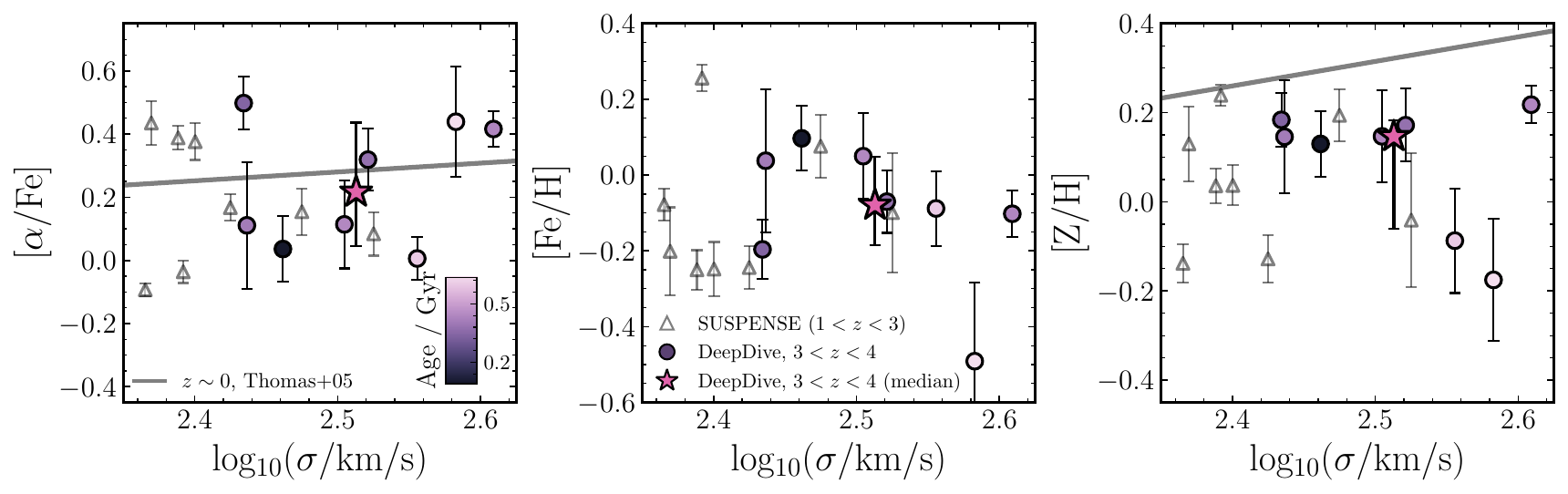}
    \caption{$\alpha$-enhancement, iron abundance ([Fe/H)] and total metallicity ([Z/H]) versus stellar velocity dispersion determined from \alfa. The pink star in each subplot represents the median [$\alpha$/Fe], [Fe/H] and [Z/H] of the \textit{DeepDive} sample, as in Fig. \ref{fig:afe_masstform}. For comparison, we also show the values we derive for the SUSPENSE galaxies at $1<z<3$ (grey triangles). Compared to the $z \sim 0$ relation, [Z/H] values are much lower, though no evolution is evident for [$\alpha$/Fe].}
    \label{fig:veldisp}
\end{figure*}

\subsection{Quenching via AGN/starburst-driven outflows}\label{sec:SFtimescalesvalpha}

Active galactic nuclei feedback is intricately connected to the evolution of galaxies, playing a crucial role in regulating star formation and quenching. Recent observations have shown evidence for (radiatively inefficient) feedback in the form of jets or winds \citep[][]{mcnamara2000,fabian2012}, which can prevent cooling flows into galaxies, and subsequently halt star formation. However, in the radiatively efficient regime, high-energy accretion can heat and expel gas on short timescales, which can also significantly affect star formation in galaxies, and particularly at early times (\citealt{choi2020}). 
At high redshift, massive galaxies are expected to produce outflows that are more effective in ejecting material, due to the higher gas fractions involved in AGN feedback \citep[see][]{leethochawalit,beverage2025,cheng2025}. Massive quiescent galaxies at $z\sim 1$ have been found to have higher mass-loading factors than nearby galaxies at fixed stellar mass, suggesting that galaxies at higher redshift experience more enhanced outflows during star formation, leading to earlier quenching \citep[see][]{leethochawalit,zhuang2023,gountanis2025}.
Indeed, recent theoretical models suggests that radiatively efficient AGN winds at high accretion rates can effectively quench massive galaxies significantly earlier ($z\sim 6$) than when only assuming jet-mode feedback \citep[see][for AGN wind model, {\scshape mistral}]{farcy2025}, in good agreement with the quenching times found for the {\it DeepDive} sample. 

At low redshifts, early-type galaxies show a more fundamental relationship between metallicities (and element abundances) and dynamical mass, rather than stellar mass \citep[see][]{gallazzi2006}. Future work will explore these relationships between chemical abundances with dynamical mass. More recent work, however, suggests that a stronger, fundamental relation exists between metallicity and central velocity dispersion in quiescent galaxies, attributed to the integrated effect of AGN feedback in massive galaxies \citep{cappellari2016,li2018,baker2024}.   
We therefore show the relationship between [$\alpha$/Fe] and velocity dispersion (obtained from \alfa) in Fig. \ref{fig:veldisp}. We find no clear correlation between [$\alpha$/Fe] and $\sigma_{\star}$. Similarly, for [Fe/H] and [Z/H] (presented in the middle and right-hand panels of Fig. \ref{fig:veldisp} respectively), the correlations are flat with velocity dispersion. 

This may not be wholly surprising due to the small dynamic range probed by our sample (0.5-dex and 0.3-dex ranges in stellar mass and velocity dispersion) and the large scatter in $\alpha$-enhancement and metallicities. In general, metallicity increases with $\sigma_{\star}$, at higher and lower redshifts, though this appears to fall off at the highest stellar masses and velocity dispersions \citep[e.g.][]{carnall2022,gallazzi2025}. This logically indicates that we are tracing the highest mass objects, where the mass-metallicity relation plateaus with the highest velocity dispersions. Findings suggest that galaxies with lower dynamical masses, and hence shallower gravitational potential wells, clear metals out of the galaxy more easily from winds driven by Type II supernovae early in their lifetimes, without strongly affecting the gas needed for subsequent star formation episodes \citep[][]{trager2000,gallazzi2006}. Consequently, as SNe-driven winds can occur at early times, more $\alpha$ elements could be ejected compared to Fe-peak elements before the onset of Type Ia supernovae in galaxies. 
At higher redshift, it may also be possible that feedback from AGN could drive strong outflows that are efficient in clearing metals from the galaxy's ISM \citep[see e.g.][]{cheng2025}. Recently, a number of high-redshift massive galaxies demonstrate evidence for strong outflows of neutral gas (e.g. Na D, Ca II), as well as high-ionization lines (e.g. [O \textsc{iii}], [Ne \textsc{v}], [N \textsc{ii}], [S \textsc{ii}]) and broadened H$\alpha$ emission \citep[see][]{belli_2024,park2024_outflows,deugenio2024,davies_2024,valentino_2025,wu2025,taylor2026}. A Several galaxies in our {\it DeepDive} sample display [O\textsc{iii}] and H$\alpha$ emission in their spectra; combined with the low metallicities derived for some of these galaxies, it is plausible that strong AGN feedback may be able to expel metals out of the galaxy. Given that AGN feedback duty cycles are thought to operate on 1-10 Myr timescales \citep{harrison2018}, it is likely that there may have been a series of these events that halted ongoing star formation. 
However, it is especially difficult to directly link current AGN activity to the primary quenching event of MQGs, especially at these early times. Instead, bursts of AGN activity may only work to \textit{maintain} quiescence by driving gas outflows or preventing gas cooling \citep[see][]{almaini2025}. The longer formation and quenching timescales observed for a few of the galaxies (e.g. DD-170, DD-80) in our sample may hint at evidence for quenching via low-energy outflows or preventative feedback mechanisms. Recent work by P. Zhu et al. in prep. shows that 13/23 galaxies in the total \textit{DeepDive} sample (including secondary targets) demonstrate (tentative) evidence for neutral sodium outflows. However, the outflow rates corresponding to these features are not high enough to enable the outflows to escape the galaxy, implying a fountain-like mechanism in which neutral gas outflows are recycled, rather than contributing to the main quenching event. This suggests that metal ejection via this pathway may not be possible, and the lower values of [$\alpha$/Fe] found in the \textit{DeepDive} sample may instead likely be due to earlier enrichment from Type Ia supernovae.
% \textbf{Do the low [$\alpha$/Fe] gals have high [Fe/H], cos that would also suggest the earlier enrichment pathway i guess. 
% }
Finally, the lack of evolution in metallicities between $z\sim 0$ and $z\sim 4$, as demonstrated in Fig. \ref{fig:massmet}, may indicate that after AGN feedback has halted in-situ star-formation, this prevents any further enrichment of the gas, allowing these galaxies to evolve without any significant changes in their overall metallicity to $z\sim 0$. This may also hint that mergers either had no or little role to play between these two epochs or, if they did, then they may have instead involved galaxies with similar metal contents. 

\subsubsection{The delay-time distribution of Type Ia Supernovae}
Another possible scenario to account for the broad range of [$\alpha$/Fe] values in our sample is the range of star-formation histories derived from our \pipes\ fitting. Longer derived SFH timescales indicate more enrichment by Type Ia supernovae. Although the galaxies in our sample are young, the timescales of formation and quenching vary from 100 to 400 Myr; earlier enrichment ($<100$\,Myr) via SNIa can therefore have significant impacts on the chemical abundances derived for samples at high redshift. It us therefore important to account for the delay time distribution (DTD) of SNIa. This is essentially the time-dependent rate of SNIa explosions after a burst of star formation, and has been found to follow a power law of the form DTD($\tau) \propto t^{-1}$ \citep[see e.g.][]{delucia2014}. 

Chemical enrichment from SNIa can begin as early as 40-100 Myr, increasing the Fe abundance and decreasing the overall [$\alpha$/Fe] ratios much earlier. This would have a strong effect on the diversity of $\alpha$-enhancements as it would lower [$\alpha$/Fe] ratios, due to more Fe and Fe-peak elements being produced at earlier times, given that the duration of the SF from (e.g. Fig. 3, table with sf timescales) is long enough for Fe-enhancement via Type Ia SNe. Recent theoretical work has found that using a universal DTD results in a broad range of $\alpha$-enhancements for massive galaxies (using [O/Fe] as a proxy for $\alpha$-enhancement). Similarly, simple galaxy chemical evolution (GCE) models show that by lowering the DTD of SNIa to 0.05 Gyr, the  [Mg/Fe] (as a proxy for $\alpha$-enhancement) drops by $0.05$ dex due to the earlier enrichment to the ISM by Fe \citep[][]{gountanis2025}. Importantly, the minimum possible delay time for a galaxy to be enriched by SNIa is $\sim 40$\,Myr, corresponding to the minimum amount of time required to form white dwarf stars in a galaxy. Within our {\it DeepDive} sample, the range of timescales and metallicities may point to a similar earlier enrichment from SNIa, lowering the overall $\alpha$-enhancement of the sample. 
% \textbf{another thing we haven't really gone into detail is the IMF}. 
Future follow-up studies using a range of theoretical models/simulations combining the effect of earlier enrichment of SNIa and varying SFHs on the chemical abundances in high-redshift massive quiescent galaxies may shed light on the distribution of [$\alpha$/Fe] values found in this work. 

\subsection{Environment}

A high number of massive quiescent galaxies discovered at high redshift reside in dense environments \citep[][]{degraaff2024,carnall2024,kakimotor2024, mcconachie_2025,ito2025_cosmicvine}. Dense environments are thought to strongly impact formation, and should be taken into consideration when studying these high-redshift systems. Environment may have an additional impact on the $\alpha$-enhancement of massive galaxies. However, due to the formation timescales related to $\alpha$-element production in galaxies, this is thought to be a somewhat weak trend based on observations at lower redshift \citep[][]{watson2022} and simulations of $z>3$ MQGs \citep[][]{delucia2025}. At $z \sim 0.1$, the [$\alpha$/Fe] of massive quiescent galaxies is relatively unchanged between different galaxy environments, though environment does affect the stellar mass and total metallicity \citep[][]{gallazzi2021}. A number of galaxies in the \textit{DeepDive} sample have been reported to reside in over-dense environments as part of larger galaxy groups. One of the galaxies in our \textit{DeepDive} sample with low [$\alpha$/Fe] (DD-196) resides in the \textit{Cosmic Vine}, \citep[see also][]{jin2024,ito2025_cosmicvine}, a dense structure at $z = 3.44$, which may have a potential impact on its formation and quenching mechanisms, though this is hard to quantify without simulations. Another galaxy in our sample (DD-134) is part of the Jekyll and Hyde galaxy pair \citep[see][]{schreiber2018}, which has been identified to be in a larger group of MQGs from filler targets of \textit{DeepDive} (Kakimoto et al., in prep.). By contrast, this galaxy has high [$\alpha$/Fe], suggesting that the effects of large-scale environment on galaxy chemical enrichment and star-formation histories are complicated and require further observational and theoretical analysis.

\subsection{Modeling challenges \& caveats}

In this work, we have obtained star-formation histories, stellar masses, ages, metallicities from sophisticated spectro-photometric fitting using SED code \pipes\, as well as [$\alpha$/Fe] and [Fe/H] abundances from chemical abundance modeling using \alfa.  We find that galaxies at high-redshift formed early and fast, and demonstrate evidence for $\alpha$-enhancement. Our galaxies  display strong scatter in total metallicity and [$\alpha$/Fe] abundances, possibly explained by quenching processes that generate strong outflows causing efficient gas removal at early times. 

Although our results are in good agreement with expected trends for massive quiescent galaxies at $z>3$ -- and with those from lower-redshift studies -- it is important to acknowledge that the stellar population models currently being used to describe the early quiescent galaxy population are a significant bottle-neck to our understanding of massive galaxy evolution; most do not cover the young stellar population ages required at these early time, and do not account for their varying star-formation histories. 
As discussed, it is difficult to compare and draw robust conclusions between individual galaxy abundance measurements due to the methodological and model differences across studies, although qualitatively, results are in agreement. Our [$\alpha$/Fe] values should trace [Mg/Fe] \citep[see][]{knowles2023}, and broadly, our values for [$\alpha$/Fe] in SUSPENSE  and [Mg/Fe] \citep[][]{beverage2025} appear to be in good agreement, though with large uncertainties on both measurements. Considering that both measurements are derived using different sets of models which employ different isochrones, stellar spectra, solar abundance references \citep[see Table 2 in ][]{knowles2023} and the \cite{conroy_2018} models incorporating individual element response functions (so each element abundance is allowed to vary freely, unlike \citealt{knowles2023} models where all $\alpha$ elements are tied to [Mg/Fe]), these results are reassuring and suggest that our higher redshift results are robust against our model uncertainties.

Previous versions of $\alpha$-enhanced models demonstrate large differences between [Mg/Fe], [$\alpha$/Fe] and [Fe/H] values across models; \cite{2jafariyazani_2025} find $\sim 0.3$ dex between [$\alpha$/Fe] from \cite{vazdekis2015} and [Mg/Fe] from \cite{conroy_2018}, with the same difference in [Fe/H] between both models. In this paper, we find that although there are large uncertainties on individual galaxy values, our derived [$\alpha$/Fe] values for the SUSPENSE galaxies follow the [Mg/Fe] values derived by \cite{beverage2025}. 

More recently, $\alpha$-enhanced stellar population synthesis models such as $\alpha$-MC based on the $\alpha$-enhanced MIST isochrones and C3K spectral libraries have been developed for smooth integration into FSPS for use in SED codes such as \texttt{Prospector} and \pipes. These models will have substantial impacts on SED fitting of young, high-redshift quiescent galaxies, allowing us to derive more robust stellar masses, ages and star-formation histories as well as more accurate chemical abundances. Future work employing these different sets of models on the {\it DeepDive} galaxies will also highlight the model dependencies on chemical abundance and timescale measurements.

\section{Summary} \label{sec:summary}

In this work, we use deep, medium-resolution spectroscopy from the \textit{JWST} Cycle 2 \textit{DeepDive} program to investigate the star-formation histories of ten massive quiescent galaxies at $3<z<4$. Using sophisticated spectro-photometric fitting code \pipes\ we infer robust stellar masses, metallicities, ages and star-formation timescales. For the first time at these high redshifts, we measure $\alpha$-enhancement by employing \alfa\ \citep[][]{beverage2025} to obtain [$\alpha$/Fe] and [Fe/H] abundances. The main conclusions of this paper are summarized as follows: 

\renewcommand{\theenumi}{(\roman{enumi})}%
\begin{enumerate}
\setlength\itemsep{0.5em}
    \item We find that \textit{DeepDive} galaxies formed and quenched early in the Universe, deriving an early median formation time for our sample of $t_{\mathrm{form}}/\mathrm{Gyr} = 1.24 \pm 0.06$, corresponding to $z_{\mathrm{form}} \simeq 4.8$. These galaxies possess young mass-weighted ages, with a median age corresponding to $0.44\pm 0.06$\,Gyr, in good agreement with values derived for massive quiescent galaxies at similar redshifts \citep[e.g.][]{carnall2024,antwi-danso_2025,nanayakkara_2025}. 
   
    \item 
    The {\it DeepDive} galaxies display a range of formation and quenching timescales, with evidence for both long ($>200$\,Myr) and short ($<100$\,Myr) timescales, consistent with recent literature results for massive quiescent galaxies at $2<z<5$ \citep[e.g.][]{baker2925flamingojades,farcy2025}. 
    
    \item  All SFH parameterizations appear to produce formation and quenching times in good agreement with each other, however, the double-power law and bursty SFH priors produce star-formation \textit{timescales} systematically shorter (by $\simeq 70$\,Myr) than our fiducial SFH fits, most likely due to the forced late single burst for both of these models. 
    \item The \textit{DeepDive} galaxies are $\alpha$-enhanced on average, with a median value of [$\alpha$/Fe] = $0.22^{+0.22}_{-0.17}$. The [Fe/H] values we obtain are sub-solar, with an average of [Fe/H]=$-0.08^{+0.13}_{-0.11}$, in good agreement with results at $z<3$. Taken in the context of lower redshift studies, this points to the {\it DeepDive} sample forming earlier and faster than their lower redshift counterparts at fixed stellar mass, a direct consequence of the downsizing scenario. 
    \item Our results demonstrate a wide range in [$\alpha$/Fe] values ($\sigma = 0.3$ dex) with some galaxies exhibiting lower ($\sim 0.1$-dex) [$\alpha$/Fe] ratios than expected at these redshifts. The timescales associated with these low measurements suggest longer-than-physical timescales ($>4$ Gyr) at these early times. This may point to an earlier onset of Type Ia SNe enrichment (earlier DTD), or metals preferentially removed early in outflows from strong winds driven by Type II SNe or AGN. Future work accounting for these scenarios will shed light on the distribution of [$\alpha$/Fe] values at $z>3$. 
\end{enumerate}
In this paper, we find that although we have shown that massive galaxies at high redshift are $\alpha$-enhanced, alone, [$\alpha$/Fe] obtained from SSP models traces the formation timescales weakly at these early times. Future work must combine individual element abundance measurements with star-formation histories using $\alpha$-enhanced stellar population models integrated into SED fitting codes. In addition to $\alpha$-enhancement, the physics of later stellar evolutionary stages (a.g. from AGB stars) must be better incorporated into our models as we obtain better data from \textit{JWST} probing longer wavelengths ($>7500$\AA) previously unavailable with \textit{HST} and ground-based studies \citep[see e.g.,][]{lu2025}, as these can have significant impact on the initial mass function of massive galaxies at high redshift. 
Similarly, the contribution of AGN contamination in most high-redshift massive quiescent galaxies detected so far calls for better integration of AGN models in SED fitting codes. 

Overall the \textit{DeepDive} galaxies represent a unique sample to study an important time in cosmic evolutionary history. Expanding this dataset with the addition of lower and higher redshift quiescent samples (as done with the SUSPENSE galaxies in this work) will allow us to trace the evolution of $\alpha$-enhancement, total stellar metallicities and star-formation histories over a wide range of cosmic time. Access to models with younger stellar populations such as those used in this paper \citep{knowles2023}, and those which will be used in future work \citep[e.g. $\alpha$-MC in FSPS,][]{park2024a}, which can accurately model the diversity in metallicities, ages, and star-formation histories of massive quiescent galaxies at earlier times. This has afforded us the capability to explore parameter spaces previously unavailable with solar-scaled models in SED fitting codes.

\begin{acknowledgements}
MLH would like to thank a number of colleagues for insightful discussions that helped improve the quality of this paper: L. Barrufet, S. Berek, R. Cochrane, K. Iyer, S. Manning, M. de los Reyes and T. M. Stanton. 
This work is based in part on observations made with the NASA/ESA/CSA James Webb Space Telescope and the NASA/ESA Hubble Space Telescope obtained from the Space Telescope Science Institute, which is operated by the Association of Universities for Research in Astronomy, Inc., under NASA contract NAS 5–26555. The data were obtained from the Mikulski Archive for Space Telescopes at the Space Telescope Science Institute, which is operated by the Association of Universities for Research in Astronomy, Inc., under NASA contract NAS 5-03127 for JWST. 
KEW and MLH gratefully acknowledge financial support for program JWST-GO-03567, provided by NASA through grants from the Space Telescope Science Institute, which is operated by the Associations of Universities for Research in Astronomy, Incorporated, under NASA contract NAS 5-03127.  Some of the data products presented herein were retrieved from the Dawn JWST Archive (DJA). DJA is an initiative of the Cosmic Dawn Center, which is funded by the Danish National Research Foundation under grant No. 140.
The Dunlap Institute is funded through an endowment established by the David Dunlap family and the University of Toronto. 
TK acknowledges support from JSPS grant 25KJ1331.
FV, KI, and PZ acknowledge support from the Independent Research Fund Denmark (DFF) under grant 3120-00043B.
The Cosmic Dawn Center (DAWN) is funded by the Danish National Research Foundation under grant DNRF140. WMB gratefully acknowledges support from DARK via the DARK fellowship. This work was supported by a research grant (VIL54489) from VILLUM FONDEN.
JRW acknowledges that support for this work was provided by The Brinson Foundation through a Brinson Prize Fellowship grant.

\end{acknowledgements}

\bibliographystyle{aasjournalv7}
\bibliography{deepdive_bib}

\appendix

\begin{figure*}
\centering
    \includegraphics[width = 0.75\linewidth]{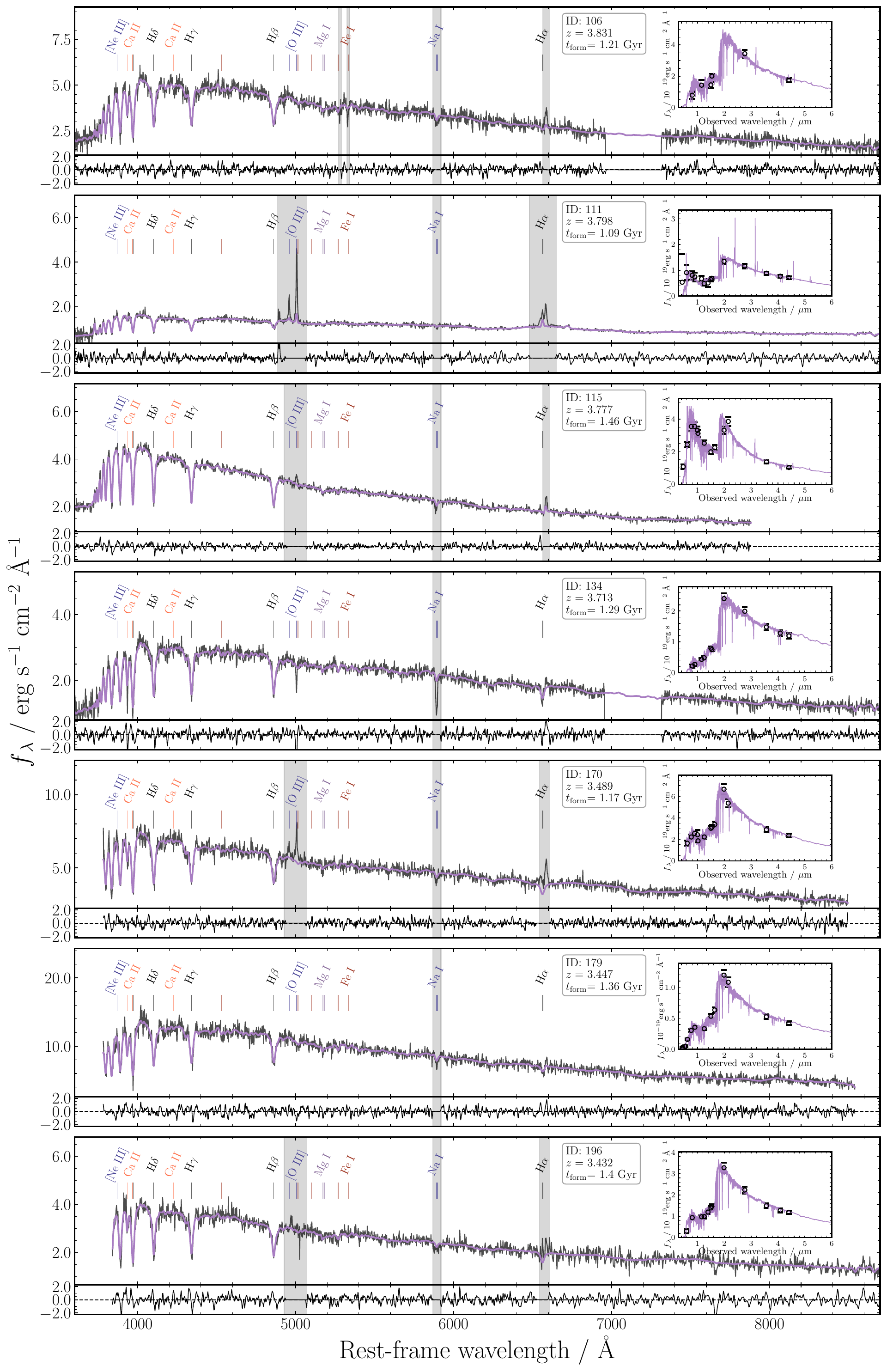}
    \caption{The same as Fig. \ref{fig:allgals_spec} for the rest of the galaxies in our main {\it DeepDive} sample. }
\end{figure*}

\end{document}